\documentclass[prc,nofootinbib,amsmath,amssymb,superscriptaddress,groupedaddress,showpacs]{revtex4-1}
\usepackage{epsfig}
\usepackage{dcolumn}
\usepackage{bm}
\usepackage{amssymb}
\usepackage{amsmath}

\newcommand{\be}{\begin{equation}}
\newcommand{\ee}{\end{equation}}
\newcommand{\bea}{\begin{eqnarray}}
\newcommand{\eea}{\end{eqnarray}}

\begin{document}
%\title{Predictions for high $p_T$ flow harmonic cumulants at LHC $\sqrt{s_{NN}}=5.02$ TeV}
\title{Cumulants and nonlinear response of high $p_T$ harmonic flow at $\sqrt{s_{NN}}=5.02$ TeV}

\author{Jacquelyn Noronha-Hostler$^a$}
\author{Barbara Betz$^b$}
\author{Miklos Gyulassy$^{c,d,e}$}

\author{Matthew Luzum$^f$}
\author{Jorge Noronha$^f$}

\author{Israel Portillo$^a$}
\author{Claudia Ratti$^a$}
\affiliation{\small{\it $^a$ Department of Physics, University of Houston, Houston, TX 77204, USA}}
\affiliation{\small{\it $^b$ Helene-Wessel-Str.\ 12, 53125 Bonn, Germany}}
\affiliation{\small{\it $^c$ Nuclear Science Division, Lawrence Berkeley National Laboratory, Berkeley, CA 94720, USA}}
\affiliation{\small{\it $^d$ Pupin Lab MS-5202, Department of Physics, Columbia University, New York, NY 10027, USA}}
\affiliation{\small{\it $^e$ Institute of Particle Physics, Central China Normal University, Wuhan, China}}
\affiliation{\small{\it $^f$ Instituto de F\'{\i}sica, Universidade de S\~{a}o Paulo, C.P. 66318,
05315-970 S\~{a}o Paulo, SP, Brazil}}

%Authors in alphabetical order

\date{\today}
\begin{abstract}
Event-by-event fluctuations caused by quantum mechanical fluctuations in the wave function of colliding nuclei in ultrarelativistic heavy ion collisions were recently shown to be necessary for the simultaneous description of $R_{AA}$ as well as the elliptic and  triangular flow harmonics at high $p_T$ in PbPb collisions at the Large Hadron Collider. In fact, the presence of a finite triangular flow as well as cumulants of the flow harmonic distribution that differ from the mean are only possible when these event-by-event fluctuations are considered. In this paper we combine event-by-event viscous hydrodynamics and jet quenching to make predictions for high $p_T$ $R_{AA}$, $v_2\{2\}$, $v_3\{2\}$, and $v_2\{4\}$ in PbPb collisions at $\sqrt{s_{NN}}=5.02$ TeV. With an order of magnitude larger statistics we find that high $p_T$ elliptic flow does not scale linearly with the soft elliptical flow, as originally thought, but has deviations from perfectly linear scaling. A new experimental observable, which involves the difference between the ratio of harmonic flow cumulants at high and low $p_T$, is proposed to investigate the fluctuations of high $p_T$ flow harmonics and measure this nonlinear response. By varying the path length dependence of the energy loss and the viscosity of the evolving medium we find that $R_{AA}(p_T)$ and $v_2\{2\}(p_T)$ strongly depend on the choice for the path length dependence of the energy loss, which can be constrained using the new LHC run 2 data.
\end{abstract}

\maketitle

\section{Introduction}

In recent years, the cumulants of low $p_T$ azimuthal flow harmonic distributions measured in ultrarelativistic heavy ion collisions have been used to attest to the collective behavior of the Quark-Gluon Plasma (QGP) and its description using event-by-event viscous hydrodynamics (for  reviews, see \cite{Heinz:2013th,Luzum:2013yya,deSouza:2015ena}). For PbPb collisions at the LHC it was found that there is a clear separation between the 2- and 4-particle elliptic flow cumulants, $v_2\{2\}$ and $v_2\{4\}$, respectively, followed by an approximate convergence of higher order cumulants, i.e., $v_2\{2\} > v_2\{4\} \approx v_2\{6\} \approx v_2\{8\}$ \cite{Chatrchyan:2013kba,Abelev:2014mda,Aad:2014vba}. In pPb collisions, where the system formed is considerably smaller, the same behavior for the multiparticle flow cumulants is observed \cite{Abelev:2014mda,Khachatryan:2015waa}. Also, quite strikingly, a similar pattern involving the cumulants of soft anisotropic flow coefficients appears in high multiplicity events in pp collisions at the LHC \cite{Aad:2015gqa,Khachatryan:2016txc} though in this case $ v_2\{4\}$ is closer to $v_2\{2\}$ than it is in larger systems \cite{Khachatryan:2016txc}.

A significant body of research has been developed for studying how initial state fluctuations translate into the final flow harmonics at low $p_T$. Small scale subnucleon fluctuations were found to have negligible effect on the lowest order flow harmonics \cite{Noronha-Hostler:2015coa}, whereas some sensitivity can be found for sub-leading modes \cite{Mazeliauskas:2015efa}. In fact, the global shape of the initial condition dominates the description of the flow harmonics at low $p_T$.  For elliptical and triangular flows, $v_2$ and $v_3$, respectively, there is a primarily linear mapping between the eccentricity of the initial state, $\varepsilon_2$, $\varepsilon_3$ and the final $v_2$, $v_3$, i.e., $v_2 \sim \varepsilon_2$ and $v_3 \sim \varepsilon_3$ when the QGP is modeled as a nearly perfect fluid \cite{Teaney:2010vd,Qiu:2011iv,Gardim:2011xv,Teaney:2012ke,Niemi:2012aj,Gardim:2014tya} (nonlinear corrections only become relevant in this case for peripheral collisions \cite{Niemi:2015qia,Noronha-Hostler:2015dbi}). On the other hand, higher order flow harmonics exhibit nonlinear response via mode mixing \cite{Qiu:2011iv,Gardim:2011xv,Teaney:2012ke,Niemi:2012aj,Gardim:2014tya}. Additionally, deviations between higher order cumulants at low $p_T$ may be attributed  to the skewness of the initial eccentricity fluctuations \cite{Gronqvist:2016hym,Giacalone:2016eyu}. 

Overall, the mapping between initial state fluctuations and the final flow harmonics in the soft sector has been very successful to the point that event-by-event viscous hydrodynamics \cite{Niemi:2015voa,Noronha-Hostler:2015uye} was able to accurately predict an increase on the order of a few percent in the flow harmonics at LHC when the collision energy was raised from $\sqrt{s_{NN}}=2.76$ TeV to $\sqrt{s_{NN}}=5.02$ TeV \cite{Adam:2016izf}. This gives support to the current understanding that the initial spatial anisotropies generated by quantum fluctuations in the wave function of the incident nuclei, when combined with event-by-event hydrodynamic simulations for the strongly coupled nearly perfect QGP fluid, can account for the experimentally observed pattern of low $p_T$ azimuthal flow harmonics.   

Meanwhile, theoretical understanding of the connection between initial state fluctuations and the experimentally observed flow harmonics at high $p_T$ is still in its infancy. The tomographic aspects of the standard jet quenching-related observables, the nuclear suppression factor $R_{AA}$ and its azimuthal Fourier components, make them in principle sensitive to the details of the many aspects of our current multilayered description of the bulk QGP evolution such as: the choice for the initial conditions, the dimensionality of the hydrodynamical evolution (i.e., 2+1 or full 3+1 hydrodynamic simulations), the temperature dependence of the transport coefficients \cite{NoronhaHostler:2008ju,NoronhaHostler:2012ug,Denicol:2012cn,Finazzo:2014cna,Rougemont:2015ona,Noronha-Hostler:2015qmd} and its connection with the QGP equation of state \cite{Borsanyi:2013bia,Bazavov:2014pvz}, the later stages of hadronic evolution after freeze-out and etc. A systematic study of the many phenomenological parameters currently involved in the hydrodynamic description of the QGP at low $p_T$ can be found in \cite{Bernhard:2016tnd}.

An investigation of the influence of these many factors on observables in the hard sector can be carried out by coupling jet tomography models with full event-by-event viscous hydrodynamics, as done for the first time in \cite{Noronha-Hostler:2016eow}. There, it was pointed out that the calculation of high $p_T$ azimuthal coefficients, which are experimentally defined via a nontrivial correlation between soft and hard particles over many events, necessarily requires the use of event-by-event hydrodynamics. In fact, by including the hydrodynamic evolution \cite{Noronha-Hostler:2013gga,Noronha-Hostler:2014dqa} of the initial stage energy density fluctuations in the soft sector and its influence in the hard sector using a simplified jet energy loss model \cite{Betz:2011tu,Betz:2012qq,Betz:2014cza}, a simultaneous description of high $p_T$ $R_{AA}$, $v_2\{2\}$, and $v_3\{2\}$\footnote{Note that $v_3\{2\}$ arises only in the presenece of event by event fluctuations \cite{Alver:2010gr}. } at LHC $\sqrt{s_{NN}}=2.76$ TeV was obtained for the first time in Ref.\ \cite{Noronha-Hostler:2016eow}.   A further test of this is to make predictions for the $R_{AA}$ and flow harmonics across different collision energies, centralities, and other types of collisions (e.g. pPb).

In this paper predictions are made for $v_2\{2\}(p_T>10 {\rm \, GeV})$, $v_2\{4\}(p_T>10 {\rm \, GeV})$, and  $v_3\{2\}(p_T>10 {\rm \,GeV})$ at LHC $\sqrt{s_{NN}}=5.02$ TeV for PbPb collisions using event-by-event relativistic hydrodynamics (modeled via the v-USPhydro code \cite{Noronha-Hostler:2013gga,Noronha-Hostler:2014dqa}) and jet tomography (the BBMG model \cite{Betz:2011tu,Betz:2012qq,Betz:2014cza}). Special care is taken in the theoretical evaluation of these quantities to reproduce the technical procedures used in the experiment, such as the multiplicity weighing process involved in the calculation of the cumulants. We investigate the sensitivity of these observables to the choice of the path length dependence of the energy loss, i.e., $dE/dL \sim L$ or $dE/dL \sim L^2$, the shear viscosity to entropy density ratio, $\eta/s$, of the hydrodynamic background, and the jet decoupling parameter (a value of the temperature in the hadronic phase below which energy loss is assumed to vanish). We find that the path length dependence of the energy loss plays a significant role in the calculation of $R_{AA}$ and  multiparticle cumulants of high $p_T$ elliptic flow for all centralities while viscosity becomes more relevant in peripheral collisions. On the other hand, we find that viscosity contributes to the decorrelation of soft vs. hard event plane angles. Future LHC PbPb run 2 data at $\sqrt{s_{NN}}=5.02$ TeV will be crucial to determine which type of energy loss model is preferred. 

A novel theoretical feature about high $p_T$ anisotropic flow uncovered in this work concerns the \emph{approximate} linear relationship between the event-by-event evaluated soft and hard $v_2$'s discussed in \cite{Noronha-Hostler:2016eow}. A careful analysis involving an order of magnitude more events than used in \cite{Noronha-Hostler:2016eow} reveals that the high $p_T$ $v_2^{\rm hard}$ does not scale perfectly linearly with its soft sector counterpart, $v_2^{\rm soft}$, but rather has some nonlinear scaling that produces novel results in the cumulants. This deviation from linear response stems from the tomographic nature of the jet energy loss calculations and produces, as a direct consequence, a different value for the $v_2\{4\}(p_T)/v_2\{2\}(p_T)$ ratio in the hard sector in comparison to the corresponding quantity at low $p_T$. The confirmation of this nonlinear effect could be readily verified using high $p_T$ elliptic flow cumulants from LHC PbPb run 2 data.  

This paper is organized as follows. In the next section we give the details about our jet+viscous hydrodynamics model. In Section \ref{bigsection} we discuss the importance of event-by-event fluctuations at high $p_T$ and define the hard sector observables computed in this paper. The dependence of elliptic flow at high $p_T$ with the initial state energy density eccentricities is presented in Section \ref{nonlinearity}. Predictions for LHC PbPb data at $\sqrt{s_{NN}}=5.02$ TeV are shown in Section \ref{results}. A study about the correlation between the event planes in the soft and the hard sectors is done in \ref{decorrelation}. We finish with our conclusions and outlook in \ref{final}. 

%%%%%%%%%%%%%%%%%%%%%%%%%%%%%%%%%%%%%%%%%%%%

\section{Combining event-by-event hydrodynamics with jet tomography}
\label{definemodel}

In this paper we use the same jet energy loss + event-by-event viscous hydrodynamic setup employed in \cite{Noronha-Hostler:2016eow} now to investigate the case of PbPb collisions at $\sqrt{s_{NN}}=5.02$ TeV. Viscous hydrodynamics is used to model the soft sector on an event-by-event basis and describe the flow harmonics at low $p_T$. The hydrodynamic fields for each event are then used in the jet energy loss model, which determines the nuclear modification factor and the properties of the flow harmonics in the hard sector event-by-event. The specific details of our model can be found below.

\subsection{Hydrodynamic model}

The hydrodynamic evolution of the QGP is modeled through 
event-by-event simulations performed using the 2+1 (i.e., boost invariant) viscous relativistic 
hydrodynamics v-USPhydro \cite{Noronha-Hostler:2013gga,Noronha-Hostler:2014dqa}. The equations of motion of viscous hydrodynamics, presented in \cite{Noronha-Hostler:2014dqa}, are solved using a Lagrangian algorithm called Smoothed Particle Hydrodynamics \cite{SPH,Aguiar:2000hw}. The accuracy of the code has been demonstrated in \cite{Noronha-Hostler:2014dqa} via a comparison to analytical and semi-analytical radially expanding solutions of 2nd order conformal hydrodynamics derived in \cite{Marrochio:2013wla}. 

The current version of v-USPhydro contains the leading terms in both the bulk and shear viscosity sectors, which define four transport coefficients: the shear viscosity $\eta$ and its relaxation time $\tau_\pi$ as well as the bulk viscosity $\zeta$ and its corresponding relaxation time $\tau_\Pi$. As in \cite{Noronha-Hostler:2016eow}, in our event-by-event simulations we set $\eta/s$ to be a constant and neglect effects from bulk viscosity. Effects from additional conserved currents, such as baryon number, are not taken into account.

The initial time of hydrodynamic simulations, $\tau_0$, was set to be $\tau_0=0.6$ fm (the initial shear stress tensor, $\pi^{\mu\nu}(\tau_0,x,y)$, is set to zero). We employed the lattice-based equation of state EOS S95n-v1 of Ref.\ \cite{EOS} and an isothermal Cooper-Frye \cite{CF} freezeout with freeze-out temperature $T_F = 120$ MeV. In v-USPhydro, particle decays are included (with hadronic resonances with masses up to 1.7 GeV) using an adapted version of the corresponding subroutine in the AZHYDRO code \cite{azhydro}. In this work, we use MCKLN initial conditions \cite{Drescher:2006ca} for the hydrodynamic simulations (see \cite{Noronha-Hostler:2015uye} for details about these initial conditions at $\sqrt{s_{NN}}=5.02$ TeV).

At the highest LHC energy the long time spent in the hydrodynamically expanding system is a predominant component in the increase of flow harmonics in the soft sector between LHC run 1 and run 2.  As shown in \cite{Noronha-Hostler:2015uye}, the change in eccentricities relevant for elliptic flow is only a $\Delta \varepsilon_2\sim\pm 1\%$ effect. However, holding the eccentricities constant but allowing for a longer hydrodynamical expansion, in  order to obtain a $20\%$ increase in the particle distribution, can generate as much as $6\%$ increase in $v_2$ for the most peripheral collisions (central collisions were found to be largely insensitive to this effect).  
%%%%%%%%%%%%%%%%%%%%%%%%%%%%%%%%%%%%
% Figure 1

\begin{figure}[ht]
\centering
\includegraphics[width=0.5\textwidth]{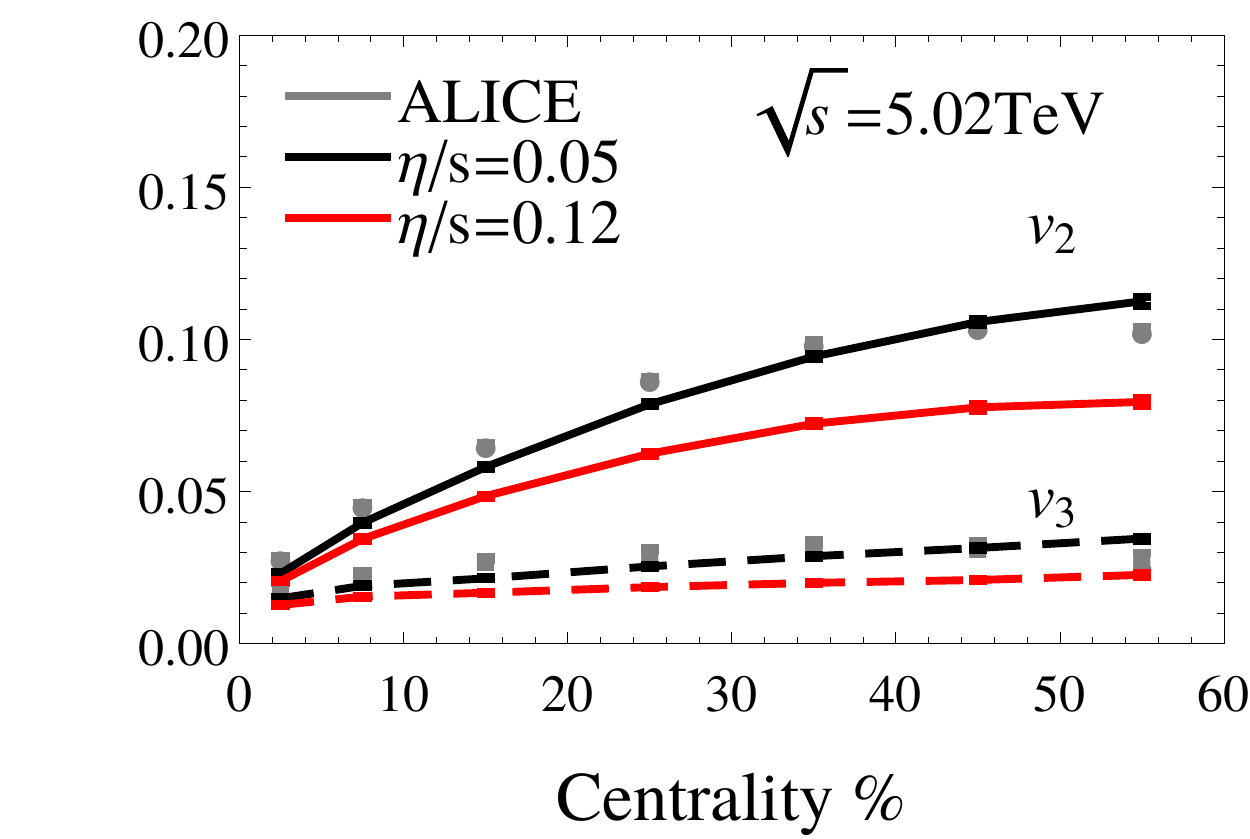}
\caption{(Color online) Model calculations for the soft sector $v_2\{2\}$ and  $v_3\{2\}$ as a function of centrality for $0.2 \leq p_T \leq 3$ GeV, computed using $\eta/s=0.05$ (black curves) and $\eta/s=0.12$ (red curves), and their comparison to ALICE $\sqrt{s_{NN}}=5.02$ TeV PbPb data \cite{Adam:2016izf}.}
\label{fig:v2v3soft}
\end{figure}

%%%%%%%%%%%%%%%%%%%%%%%%%%%%%%%%%%%%%

We show in Fig.\ \ref{fig:v2v3soft} a comparison between our model calculations\footnote{We note that the multiplicity weighing and the centrality class rebinning procedures, described in Section \ref{bigsection}, are taken into account in these calculations.} for the centrality dependence of the $p_T$-integrated 2-particle cumulants of elliptic and triangular flow, $v_2\{2\}$ and  $v_3\{2\}$, and the corresponding ALICE PbPb data at $\sqrt{s_{NN}}=5.02$ TeV \cite{Adam:2016izf}. In this plot, we used 1000 hydrodynamic events per centrality bin. A reasonable agreement with the data is found for $\eta/s=0.05$ while for $\eta/s=0.12$ the viscous suppression of the flow harmonics is not compatible with the data. 

Such a small value of $\eta/s$ is a consequence of using MCKLN initial conditions at these higher energies. In fact, at $\sqrt{s_{NN}}=5.02$ TeV one finds that MCKLN shows a $2-3\%$ decrease in  $\varepsilon_3$ while $\varepsilon_2$ is roughly constant \cite{Noronha-Hostler:2015uye}. However, ALICE measures a $4.3\%$ increase in triangular flow \cite{Adam:2016izf} so if we use the same $\eta/s=0.11$ as done for run 1 data in \cite{Noronha-Hostler:2016eow}, the low $p_T$ flow harmonics are too strongly suppressed. To compensate for this effect, here $\eta/s$ is decreased to $0.05$ to describe $\sqrt{s_{NN}}=5.02$ TeV data. One fortunate outcome of such as a small $\eta/s$ is that the reduction of sensitivity to (as yet unknown) initial state $\pi^{\mu\nu}(\tau_0,x,y)$ fluctuations. In contrast, with $\eta/s=0.2$, as for example used in \cite{Niemi:2015voa}, even small variations around the assumed initial condition for $\pi^{\mu\nu}(\tau_0,x,y)$ could result in excessively large dissipative corrections to the evolution that still need to be checked. Nevertheless, to check the sensitivity of our results with variations in $\eta/s$ we also considered the value $\eta/s=0.12$ in our calculations, as shown in Fig.\ \ref{fig:v2v3soft}.

Additionally, other effects could cause a differenc in $\eta/s$ across energies such as the fact that  the original mckln fit of $\eta/s=0.11$ was made to ATLAS data that has a different $p_T$ range than the ALICE data measured here (ATLAS starts $p_T>0.5$ GeV whereas ALICE starts with $p_T>0.2$ GeV). Furtheremore, including charm into the Equation of State appears to play a role as one continues to probe higher and higher temperatures \cite{Borsanyi:2016ksw}. 

We note that though such a small value of $\eta/s=0.05$ is below the original ``viscosity bound" previously suggested in \cite{Kovtun:2004de}, it is now understood that finite coupling and $N_c$ corrections can give values of $\eta/s$ that are indeed below $1/4\pi$ in holographic models \cite{Kats:2007mq,Brigante:2007nu,Brigante:2008gz,Buchel:2008vz}. In fact, $0.05$ is close to the bound derived in \cite{Brigante:2008gz} for a class of conformal field theories with Gauss-Bonnet gravity dual. Furthermore, a violation of the bound also appears if local spatial isotropy is broken by the presence of a strong magnetic field \cite{Critelli:2014kra,Finazzo:2016mhm}.

%%%%%%%%%%%%%%%%%%%%%%%%%%%%%%%%%%%%%%%%%%%

\subsection{Jet energy loss model} 

With all the parameters for the soft sector fixed, we now discuss the details of the jet energy loss model used in this work. In the BBMG model \cite{Betz:2014cza} the dependence of the energy loss rate with the jet energy $E$, path length $L$, temperature $T$, and energy loss fluctuations $\zeta_q$ is characterized by the parameters $(a, z, c, q)$ that appear in the following formula for the energy loss per unit length
\be
\frac{dE}{dL} = -\kappa\,%(E^2,T) 
E^a(L) \,L^z\, T^c \,\zeta_q\, \Gamma_{\rm flow}
\label{barbenergyloss}
\ee 
where $\kappa$ is the jet-medium coupling \cite{Betz:2014cza}, $c=2+z-a$, and
\be
\Gamma_{\rm flow} = \gamma \left[1-v \cos\left(\phi_{\rm jet}-\phi_{\rm flow}  \right)\right]
\label{gammaflow}
\ee 
is the flow factor defined using the local flow velocities of the medium $\vec{u}=\gamma \vec{v}$ (where $\gamma = 1/\sqrt{1-\vec{v}^{\,2}}$) \cite{Armesto:2004vz,Renk:2005ta,Baier:2006pt}. This term is important since it couples the differences in path length in the medium to the energy loss experienced by the partons. Moreover, in \eqref{gammaflow} $\phi_{\rm jet}$ is the angle defined by the propagating jet in the transverse plane while $\phi_{\rm flow}$ is the local azimuthal angle of the medium constructed using the spatial components of the hydrodynamic flow velocity. The $\kappa$ parameter in the BBMG energy loss model is completely fixed by setting the computed $\pi^0$ $R_{AA}(p_T = 10 {\rm \,GeV})\approx 0.17$. We note that in our model effects from the viscosity of the medium on the magnitude of the energy loss are highly indirect since they only appear via the temperature and flow velocity dependence of \eqref{barbenergyloss}.

Besides the ``pQCD-scenario" used in \cite{Betz:2014cza,Noronha-Hostler:2016eow} where $(a=0,z=1,c=3,q=0)$, i.e., $dE/dL \sim L$, here we also investigate the effects of a quadratic path length dependence \cite{Marquet:2009eq,Jia:2011pi,Jia:2010ee,Adare:2012wg}, i.e., $dE/dL \sim L^2$, defined by setting $(a=0,z=2,c=4,q=0)$ in \eqref{barbenergyloss}. We will see in Section \ref{results} that both the nuclear modification factor and the flow harmonics are sensitive to this choice for the path length dependence of the energy loss. 

In our model the partonic jets are distributed according to event-by-event transverse energy density profiles of the medium given by the v-USPhydro code. The jet path $\vec{x}(L)=\vec{x}_0+\hat{n}(\phi_{\rm jet})L$ from a production point $\vec{x}_0$ is perpendicular to the beam and moves in the transverse plane along the direction defined by $\phi_{\rm jet}$. Parton distributions from LO perturbative QCD calculations \cite{private} are used. Moreover, we assume that the jets do not lose energy at the points in the medium where the local temperature is smaller than a certain energy scale, which we call the jet-medium decoupling parameter, taken to be either 120 MeV or 160 MeV (below these temperatures standard fragmentation takes place). By varying this phenomenological parameter we can assess part of the uncertainties related to the complicated process of hadronization. Also, as in \cite{Noronha-Hostler:2016eow}, we use the KKP pion fragmentation functions \cite{Kniehl:2000hk,Simon:2006xt} in our calculations at high $p_T$. For more details about the BBMG model, we refer the reader to Ref.\ \cite{Betz:2014cza}.

%%%%%%%%%%%%%%%%%%%%%%%%%%%%%%%%%%%%%%%%%%%%
  
\section{The importance of event-by-event fluctuations at high transverse momentum}
\label{bigsection}

Here we discuss how the inclusion of initial state fluctuations, and their subsequent evolution using event-by-event viscous hydrodynamics, affect the theoretical description of the nuclear modification factor and also the flow harmonics at high $p_T$. This section contains many details about how to properly compute flow harmonics at high $p_T$ in a way that can be meaningfully compared to experimental data. This discussion largely extends the brief summary presented in \cite{Noronha-Hostler:2016eow} by giving explicit expressions for the cumulants of flow harmonics involving soft and hard hadrons while also providing the details about the multiplicity weighing and centrality class rebinning procedures used in experimental analyses at high $p_T$.  

The energy loss experienced by fast moving partons in the QGP has been studied over the years using the nuclear modification factor
\be
R_{AA}(p_T,\phi) = \frac{1}{\mathcal{N}}\frac{dN_{AA}/dp_T d\phi}{dN_{pp}/dp_T},
\ee 
where $dN_{AA}/dp_T$ is the particle yield (e.g., pions) per event in 
AA collisions, $dN_{pp}/dp_T$ is the proton-proton yield, $\phi$ is the azimuthal angle in the plane transverse to the beam direction, and $\mathcal{N}$ is the appropriate normalization factor (for a given AA centrality) defined in terms of the number of binary collisions \cite{Miller:2007ri} and the nucleon-nucleon inelastic cross section. We note that the boost invariance assumption made in this work restricts our calculations to the mid-rapidity region, $y=0$.

The azimuthally averaged version of the nuclear modification factor \cite{Gyulassy:1990ye,Wang:1991hta,Wang:1991xy,Vitev:2002pf}
\be
R_{AA}(p_T) =\frac{1}{2\pi} \int_0^{2\pi}d\phi\, R_{AA}(p_T,\phi)
\ee
has been found experimentally \cite{Adcox:2001jp,Adler:2002xw,Adler:2003qi,Adams:2003kv,Adams:2003im,Abelev:2012hxa,CMS:2012aa} to strongly depend on global properties of heavy ion events such as their centrality (multiplicity). In fact, in the most central AA collisions where the parton density is the largest, $R_{AA}(p_T)$ at high $p_T$ is strongly suppressed in comparison to the corresponding measurement in peripheral events. This provided an experimentally accessible way to constrain the parameters (and the assumptions) involved in the theoretical modeling of jet energy loss in the QGP including the values and temperature dependence\footnote{The analysis in \cite{Burke:2013yra} gives support to the presence of a peak in $\hat{q}/T^3$ near the crossover region, which is in agreement with non-conformal models that include non-perturbative/strong coupling behavior \cite{Liao:2008dk,Li:2014hja,Rougemont:2015wca,Xu:2015bbz}.} of the jet transport coefficient, $\hat{q}/T^3$, as discussed in detail by the JET-collaboration in Ref.\ \cite{Burke:2013yra}.

Important additional information about parton energy loss and its path length dependence in the medium can be obtained by studying the azimuthal anisotropy of high $p_T$ hadrons encoded in $R_{AA}(p_T,\phi)$ \cite{Wang:2000fq,Gyulassy:2000gk,Shuryak:2001me}. In fact, while $R_{AA}(p_T)$ can be described by many different models (see \cite{Burke:2013yra}), to obtain a simultaneous description of $R_{AA}(p_T)$ and high $p_T$ elliptic flow data has proven to be considerably more challenging (see Refs.\ \cite{Betz:2014cza,Xu:2014tda} for a discussion). 

In general, the azimuthal anisotropy of $R_{AA}(p_T,\phi)$ can be studied using its Fourier harmonics, which we call $v_n^{hard}(p_T)$, defined by the series
\be
\frac{R_{AA}(p_T,\phi)}{R_{AA}(p_T)} =  1  + 2\sum_{n=1}^\infty v_n^{hard}(p_T) \cos\left[n\phi - n\psi_n^{hard}(p_T)  \right]
\ee
where
\be
\label{definevnhard}
v_n^{hard}(p_T) = \frac{\frac{1}{2\pi}\int_0^{2\pi}d\phi\,\cos\left[n\phi-n\psi_n^{hard}(p_T)\right]\,R_{AA}(p_T,\phi)}{R_{AA}(p_T)}
\ee
and 
\be
\psi_n^{hard}(p_T) = \frac{1}{n}\tan^{-1}\left(\frac{\int_0^{2\pi}d\phi\,\sin\left(n\phi\right)\,R_{AA}(p_T,\phi)}{\int_0^{2\pi}d\phi\,\cos\left(n\phi\right)\,R_{AA}(p_T,\phi)}\right).
\ee
Previous works that investigated high $p_T$ azimuthal anisotropy in the light flavor sector, for instance \cite{Betz:2014cza,Xu:2014tda}, performed their calculations using local temperature and flow profiles from a single event-averaged background given by hydrodynamics while in \cite{Molnar:2013eqa} a kinetic theory background was used. This assumption regarding the medium evolution is not realistic given our current understanding of the QGP since it neglects the important role played by initial state fluctuations and their dynamical evolution in the calculation of flow harmonics. For instance, an immediate consequence of the inclusion of event-by-event calculations is that the jet transport parameter $\hat{q}/T^3$ possesses a complicated dependence on space and time that will be different for each hydrodynamic event.   

Apart from \cite{Noronha-Hostler:2016eow}, previous calculations of high $p_T$ flow harmonics did not include event-by-event viscous hydrodynamics and,  thus, could only consider  elliptic flow since higher harmonics such as triangular flow are identically zero in this case. As a matter of fact, as stressed in \cite{Noronha-Hostler:2016eow}, high $p_T> 10$ GeV flow coefficients such as $v_2\{2\}(p_T)$ are experimentally defined in terms of a 2-particle cumulant involving a soft and a hard hadron. This quantity is intrinsically different than the idealized $v_2^{hard}(p_T)$ in \eqref{definevnhard} as it contains the information about the jet-medium interactions encoded in the correlation between soft and hard hadrons. Similar expressions for 4-particle cumulants, e.g. $v_2\{4\}(p_T)$, involving three soft particles and one hard particle can also be computed in our framework, as it will be discussed below.

% Now the theory of vn starts ... 

In the notation used in \cite{Gardim:2011xv,Gardim:2014tya}, any flow harmonic $V_n$ can be written as a complex number composed of a magnitude $v_n$ and an angle $\psi_n$, i.e.
\begin{equation}
\label{complex}
V_n=v_n\,e^{i n \psi_n }.
\end{equation}
Such a representation is useful when one wants to write the expressions for the cumulants. In fact, the correlation between the flow harmonic coefficient taken in the integrated $p_T$ ensemble (soft particles in our case), denoted by $V_n$, with another flow harmonic at a certain (high) $p_T$, denoted by $V_n(p_T) = v_n(p_T)\,e^{i n \psi_n(p_T) }$, can be simply written as
\begin{equation}\label{eqn:v2sh}
Re\{V_n V^*_n (p_T)\}=v_n v_n(p_T) \cos \left[n\left(\psi_n-\psi_n(p_T)\right)\right].
\end{equation}
Assuming the two particles are independent, this is the probability of finding the pair in a certain azimuthal harmonic $n$.

Due to finite statistics, one must average correlations over an ensemble of events.
This is typically
done in the following manner:
\begin{itemize}
\item The events are separated by their multiplicity into $0.5\%$ centrality sub-bins
\item Within each centrality sub-bin the individual flow harmonics are calculated using multiplicity weighing in order to improve statistical error bars
\item The $0.5\%$ centrality sub-bins are then recombined into larger bins, for instance, of $5\%$ or $10\%$ once again using multiplicity weighing.  
\end{itemize}
In general, multiplicity weighing is used because events with larger multiplicity have less statistical uncertainty. As shown in \cite{Gardim:2016nrr}, it is important when including multiplicity weighing to always use small enough centrality bins because, otherwise, the multiplicity weighing can distort the final results especially in cases where ratios of cumulants of different order, such as $v_2\{4\}/v_2\{2\}$, are taken.

Due to statistical limitations, in this study we will only consider $1\%$ centrality bins and we sort by the number of participants, $N_{part}$, given by our MCKLN initial conditions.  The averaging over events within the $1\%$ multiplicity centrality bins is done as in \cite{Bilandzic:2010jr,Bilandzic:2013kga} using
\begin{equation}\label{eqn:ens}
\langle \dots \rangle = \frac{\sum_i^{events} Re\{\dots\}_i W(n_s,n_h;p_T)_i}{\sum_i^{events} W(n_s,n_h;p_T)_i},
\end{equation}
where the weight of each event $W_i$ depends on the number of soft correlated particles, $n_s$, in the experimental observable as well as on the number of hard correlated particles, $n_h$, at a given $p_T$. The weight itself is derived from the total multiplicity for integrated observables or the multiplicity within a specific $p_T$ range for differential observables.  In the language of soft vs. hard physics, for soft particles the total multiplicity $M_i$ is used while for hard particles one can use the value of $R_{AA}(p_T)_i$ at a specific point in $p_T$.  In this way, the weights read:
\begin{eqnarray}
W(2,0)_i&=&M_i (M_i-1) \label{eqn:Mw1}\\
W(4,0)_i&=& M_i (M_i-1)(M_i-2)(M_i-3)\\
W(1,1;p_T)_i&=&M_i R_{AA}(p_T)_i \\
W(3,1;p_T)_i&=& M_i (M_i-1)(M_i-2)R_{AA}(p_T)_i\,. \label{eqn:Mw4}
\end{eqnarray}
After the experimental observable is obtained in the $1\%$ centrality bins then it must be recombined into a larger bin width, once again using multiplicity weighing to recombine the bins.  

We calculate the soft-hard flow harmonic cumulants across $p_T$ using this prescription. For this paper we only consider the 2 and 4 particle cumulants:
\begin{eqnarray}
v_n\{2\}(p_T)&=&\frac{d_n\{2\}(p_T)}{\left(c_n\{2\}\right)^{1/2}}\label{eqn:cumusimp2}\\
v_n\{4\}(p_T)&=&\frac{d_n\{4\}(p_T)}{\left(-c_n\{4\}\right)^{3/4}}\label{eqn:cumusimp4}
\end{eqnarray}
where
%\begin{widetext}
\begin{eqnarray}
d_n\{2\}(p_T)&=&\frac{\sum_{j=cent_{start}}^{cent_{end}} d_{n,j}\{2\}(p_T)\sum_i^{N_{ev}^j} W(1,1;p_T)_i }{\sum_{j=cent_{start}}^{cent_{end}}\sum_i^{N_{ev}^j} W(1,1;p_T)_i }\label{eqn:Mw1}\\
c_n\{2\}&=&\frac{\sum_{j=cent_{start}}^{cent_{end}} c_{n,j}\{2\}\sum_i^{N_{ev}^j} W(2,0)_i }{\sum_{j=cent_{start}}^{cent_{end}} \sum_i^{N_{ev}^j}  W(2,0)_i}\label{cn2}\\
d_n\{4\}(p_T)&=&\frac{\sum_{j=cent_{start}}^{cent_{end}} d_{n,j}\{4\}(p_T)\sum_i^{N_{ev}^j} W(3,1;p_T)_i }{\sum_{j=cent_{start}}^{cen_{end}} \sum_i^{N_{ev}^j} W(3,1;p_T)_i }\\
c_n\{4\}&=&\frac{\sum_{j=cent_{start}}^{cent_{end}} c_{n,j}\{4\}\sum_i^{N_{ev}^j} W(4,0)_i }{\sum_{j=cent_{start}}^{cent_{end}}\sum_i^{N_{ev}^j} W(4,0)_i}\,.\label{eqn:Mw4}
\end{eqnarray}
%\end{widetext}
Here the first sum is over all the sub-bins $j$ where $cent_{start}$ is the start of the centrality class and $cent_{end}$ is the end of the centrality class  (so for $20-30\%$, $cent_{start}=20$ and $cent_{start}=30$).  The second sum is over the number of events within each $1\%$ sub-bin where $N_{ev}^j$ is the number of events in the sub-bin $j$.  The method used here is the scalar product method, which allows for an unambiguous comparison between theory and experiment \cite{Luzum:2012da} unlike the previously used event plane method \cite{eventplane}. 

Returning to Eq.\ (\ref{eqn:cumusimp2}), one can see that the 2-particle cumulant is defined in terms of $d_n\{2\}$, which itself is written in terms of the quantities $d_{n,j}\{2\}$ that include a soft and a hard particle within the sub-bin $j$ 
\begin{eqnarray}
\label{dn}
 d_{n,j}\{2\}(p_T)&=&\langle V_n V^*_n (p_T) \rangle_j\\
 &=&\langle v_n v_n(p_T) \cos \left(n\left[\psi_n-\psi_n(p_T)\right]\right) \rangle_j\label{dnj2}
\end{eqnarray}
where $\langle \dots\rangle$ is defined in Eq.\ (\ref{eqn:ens}). The normalization factor can be computed using that
\begin{eqnarray}
 c_{n,j}\{2\}&=&\langle V_n V_n^* \rangle_j\\
 &=&\langle v_n^2 \rangle_j\,.
\end{eqnarray}
This shows that the denominator of $v_n\{2\}(p_T)$ in Eq.\ (\ref{eqn:cumusimp2}) is exactly the second cumulant of the soft sector, i.e., $v_n\{2\}$. Similarly, it follows that if three soft particles are correlated with one hard particle the ensemble of flow harmonics is
\begin{eqnarray}
 d_{n,j}\{4\}(p_T)&=& 2 \langle V_n V_n^*\rangle_j \langle V_n V^*_n (p_T) \rangle_j -\langle V_n V_n^* V_n V^*_n (p_T) \rangle_j\\
 &=&2\,c_{n,j}\{2\} d_{n,j}\{2\}(p_T)  -\langle v_n^2 V_n V^*_n (p_T) \rangle_j\\
 & = &2\,c_{n,j}\{2\} d_{n,j}\{2\}(p_T)-\langle v_n^3\,v_n(p_T) \cos\left(n\left[\psi_n-\psi_n(p_T)\right]\right) \rangle_j\,.\label{dnj4}
\end{eqnarray}
The normalization factor for the 4-particle cumulant in Eqs.\ \eqref{eqn:cumusimp4} is computed using that
\begin{eqnarray}
\label{cn}
 -c_{n,j}\{4\}&=&2\langle V_n V_n^* \rangle_j^2-\langle V_n V_n^* V_n V_n^*\rangle_j\\
 &=&2(c_{n,j}\{2\})^2-\langle v_n^4 \rangle_j\,.
\end{eqnarray}
One can see that the denominator in the definition of $v_n\{4\}(p_T)$ is the cubic power of the fourth cumulant of the flow harmonic in the soft sector, $(v_n\{4\})^3$, since there are three soft particles in the numerator.  

The discussion above makes it clear that consistent comparisons of theoretical calculations of high $p_T$ flow harmonics to experimental data necessarily require the use of techniques and expertise from event-by-event viscous hydrodynamics. The expressions for the soft-hard cumulants of flow harmonics presented here are valid for any type of jet energy loss model used. Our predictions for the nuclear modification factor and the flow harmonic cumulants for PbPb collisions at $\sqrt{s_{NN}}=5.02$ TeV at the LHC will be shown in the next section.

Finally, we note that while the dynamically evolving medium affects the energy loss experienced by the jets in our model, the backreaction of this energy lost by the fast parton onto the medium is not taken into account here. This type of probe approximation, commonly used in jet quenching studies, should hold to determine the properties of the flow harmonics at sufficiently high $p_T$ (e.g., $p_T>10$ GeV).  Between the soft physics hydrodynamical regime and the high $p_T$ limit ($3\lesssim p_T\lesssim 10$ GeV)  lies a region where the influence of jets in the spacetime evolution of the QGP may be relevant. 
If part of the energy lost by jets can quickly thermalize and be distributed in the medium in a collective manner, even the bulk anisotropy of the event and their low $p_T$ flow harmonic coefficients may change \cite{Pang:2009zm,Tachibana:2014lja,Andrade:2014swa,Schulc:2014jma,Bravina:2015sda}. 
Flow measurements typically enforce rapidity gaps between measured particles, in order to suppress non-flow effects.  This also has the effect of suppressing the effect of back-reaction, which will likely be limited to rapidities near the jet.  However, there still could be some effect from an away-side jet, and in measurements without rapidity gaps, such as $v_n\{4\}(p_T)$.

%%%%%%%%%%%%%%%%%%%%%%%%%%%%%%%%%%%%%%%%%%%%

\section{Deviations from linear response}
\label{nonlinearity}

While it is by now well established that the low $p_T$ lowest order harmonic flow coefficients, such as $v_2$ and $v_3$, display an approximate linear behavior with the corresponding eccentricities $\varepsilon_2$ and $\varepsilon_3$ on an event-by-event basis for most centrality classes \cite{Gardim:2011xv,Teaney:2012ke,Niemi:2012aj,Gardim:2014tya,Niemi:2015qia,Noronha-Hostler:2015dbi}, whether or not this type of linear response also holds for harmonic flow at high $p_T$ is not known.

In Ref.\ \cite{Noronha-Hostler:2016eow}, a scatter plot of $v_2^{hard}$ (see \eqref{definevnhard}), defined in the $20 < p_T < 30$ GeV bin, versus the soft $p_T$-integrated $v_2^{soft}(0.3<p_T <3\,{\rm GeV})$ showed \emph{approximate} linear response behavior for PbPb collisions at $\sqrt{s_{NN}}=2.76$ TeV. Here we investigate this question regarding linear response of harmonic flow at high $p_T$ using two values of $\eta/s$ in the soft sector and different $p_T$ cuts in the hard sector. Also, we stress that considerably larger statistics (an order of magnitude more events than in \cite{Noronha-Hostler:2016eow}) are used in the present analysis.

\begin{figure}[ht]
\centering
\includegraphics[width=0.4\textwidth]{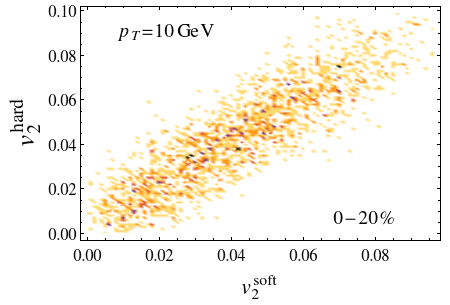}
\caption{(Color online) Event-by-event scatter plot of $v_2^{hard}$ vs.  $v^{soft}_2$ at $p_T=10$ GeV for $\eta/s=0.05$ in the $0-20\%$ centrality window.}
\label{fig:scat020}
\end{figure}

\begin{figure}[ht]
\centering
\includegraphics[width=0.4\textwidth]{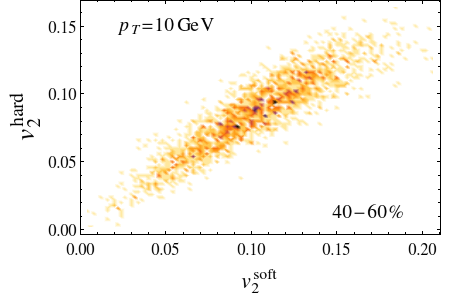}
\caption{(Color online) Event-by-event scatter plot of $v_2^{hard}$ vs.  $v^{soft}_2$ at $p_T=10$ GeV for $\eta/s=0.05$ in the $40-60\%$ centrality window.}
\label{fig:scat4060}
\end{figure}

In Figs.\ \ref{fig:scat020}-\ref{fig:scat4060} event-by-event scatter plots are shown comparing $v_2^{hard}$ vs.  $v^{soft}_2$ at $p_T=10$ GeV for $\eta/s=0.05$ in $0-20\%$ and $40-60\%$ centrality classes, respectively.  Large centrality windows are shown to improve statistics.  The approximate, yet imperfect, linear correlation is clearly visible. We note that even at low $p_T$, flow vectors at different transverse momentum are known not to be perfectly correlated \cite{Gardim:2012im}. Thus, it is not surprising to find a similar effect here, where the correlation seems to be even weaker. 

We quantify the strength of the correlation by calculating the Pearson correlation coefficient \cite{Gardim:2011xv,Gardim:2014tya} between the flow vectors $v^{soft}_2$ and $v_2^{hard}$, and also between $v^{hard}_2$ and $\varepsilon_2$.
When the two vectors are perfectly correlated this coefficient goes to 1, when they are perfectly anticorrelated it goes to -1 and when there is no linear correlation they go to zero. Here we use the symbol $Q_n$ to describe this linear correlation coefficient between two vectors, such as $v_n^{hard}$ and $v_n^{soft}$ or $v_n^{hard}$ and $\varepsilon_n$.   Written in the complex notation \eqref{complex}, this is:
\begin{equation}
\label{Qv2}
Q_n(p_T) = \frac 
{\left\langle V_n(p_T) V_n^* \right\rangle}
{\sqrt{\left\langle |V_n(p_T)|^2\right\rangle \left\langle V_n^2\right\rangle}} .
\end{equation}
The equivalent expression involving the eccentricity vector is obtained trivially by replacing $V_n\to\varepsilon_n e^{in\Phi_n}$.

%
%
%For instance, the explicit expression involving $v_2^{hard}$ and $v_2^{soft}$ is 
%\begin{equation}
%\label{Qv2}
%Q_2(p_T) = \frac 
%{\left\langle v_2^{hard}(p_T) v^{soft}_2 \cos \left[2(\Psi^{hard}_2(p_T) - \Psi_2^{soft}) \right]\right\rangle}
%{\sqrt{\left\langle (v^{hard}_2(p_T))^2\right\rangle \left\langle (v^{soft}_2)^2\right\rangle}} .
%\end{equation}
%while for $n=3$ one finds
%\begin{equation}
%\label{Qv3}
%Q_3(p_T) = \frac 
%{\left\langle v^{hard}_3(p_T) v^{soft}_3 \cos \left[3(\Psi^{hard}_3(p_T) - \Psi_3^{soft})\right] \right\rangle}
%{\sqrt{\left\langle (v^{hard}_3(p_T))^2\right\rangle \left\langle (v^{soft}_3)^2\right\rangle}} .
%\end{equation}
%In Eqs.\ (\ref{Qv2})-(\ref{Qv3}) one 
One can clearly see that the value of the Pearson coefficient is closest to one when both the magnitude of the flow harmonics and the angles are strongly correlated.  If the magnitudes were strongly correlated but the event plane angles were completely decorrelated, or vice versa, then it would still be possible to obtain zero.  

\begin{figure}[ht]
\centering
\includegraphics[width=0.4\textwidth]{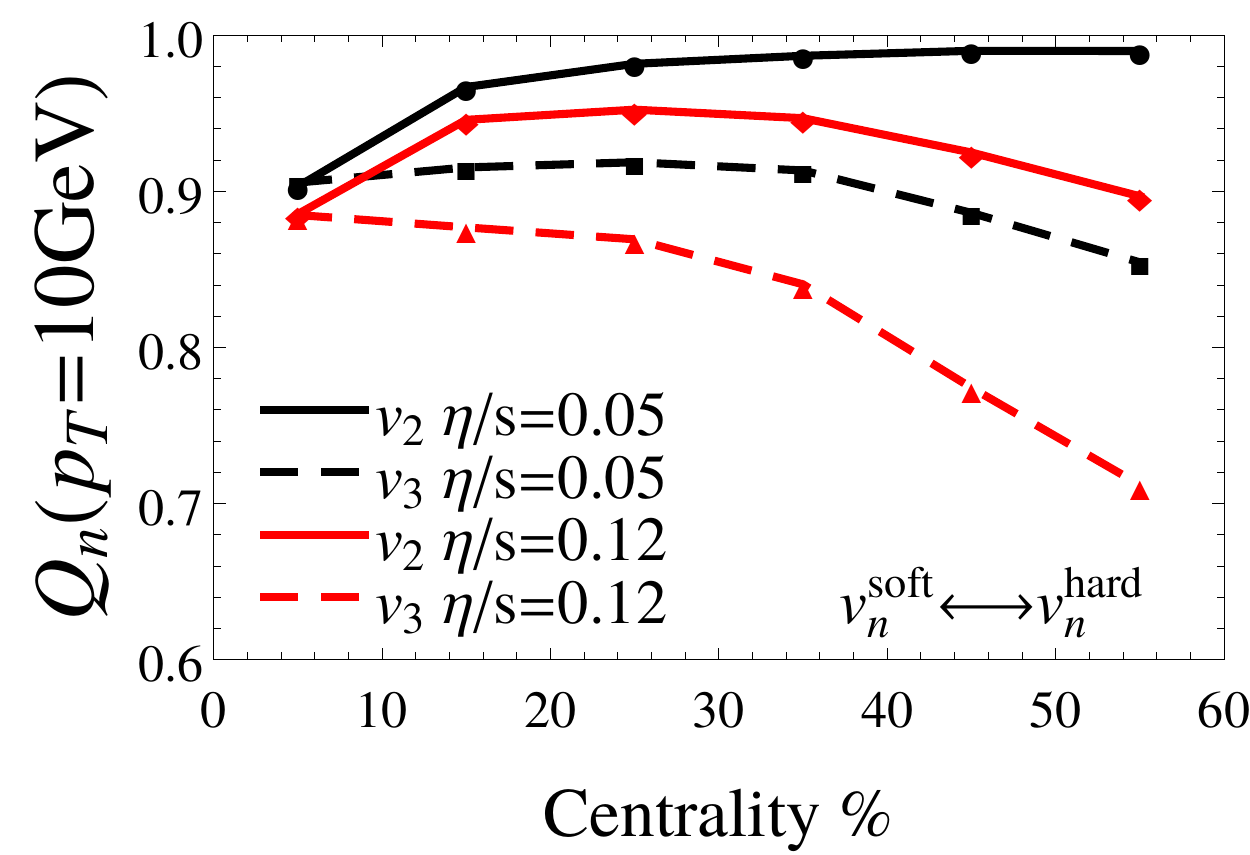}
\caption{(Color online) Pearson Coefficient, $Q_n$ between $v_n^{hard}$ and $v_n^{soft}$, across all centralities, described by Eq.\ (\ref{Qv2}) for $p_T=10$ GeV.}
\label{fig:quality}
\end{figure}
In Fig.\ \ref{fig:quality} the Pearson coefficients for elliptic and triangular flow in Eq. (\ref{Qv2}) are shown for $p_T=10$ GeV and one can see that a linear correlation between $v^{hard}_2$ and $v_2^{soft}$ is very strong for small values of the viscosity. However, for more central collisions other effects may occur since $Q_2$ clear deviates from unity and this deviation is correlated with the viscosity (larger viscosity worsens the correlation between $v^{hard}_2$ and $v_2^{soft}$).  Thus, we expect elliptic flow at high $p_T$ to display some type of nonlinear response for most central collisions and that these nonlinearities are tied to viscosity. On the other hand, Fig.\ \ref{fig:quality} shows that the hard and the soft triangular flow are not nearly as strongly correlated. The reason for this is mostly likely the decorrelation between their event plane angles, as discussed in Section \ref{decorrelation}. Also, we see that there is a significant influence of viscosity on the correlation between hard and soft triangular flow as well. We note that one rather surprising finding is that for central collisions the strength of the correlation is essentially identical for $v_2$ and $v_3$ and it may be possible that for super central collisions the linear correlation for $v_3$ is actually stronger than for $v_2$.

At higher $p_T$ the linear correlation between $v^{hard}_2$ and $v_2^{soft}$ is considerably improved, especially for central collisions.  Furthermore, viscous corrections, while having the same qualitative effect as for $p_T=10$ GeV, appear to have a smaller influence at high $p_T$.  The correlation of triangular flow worsens at higher $p_T$, which is likely due to the more decorrelated event plane angles for triangular flow at high $p_T$ seen in Section \ref{decorrelation}. 

\begin{figure}[ht]
\centering
\includegraphics[width=0.4\textwidth]{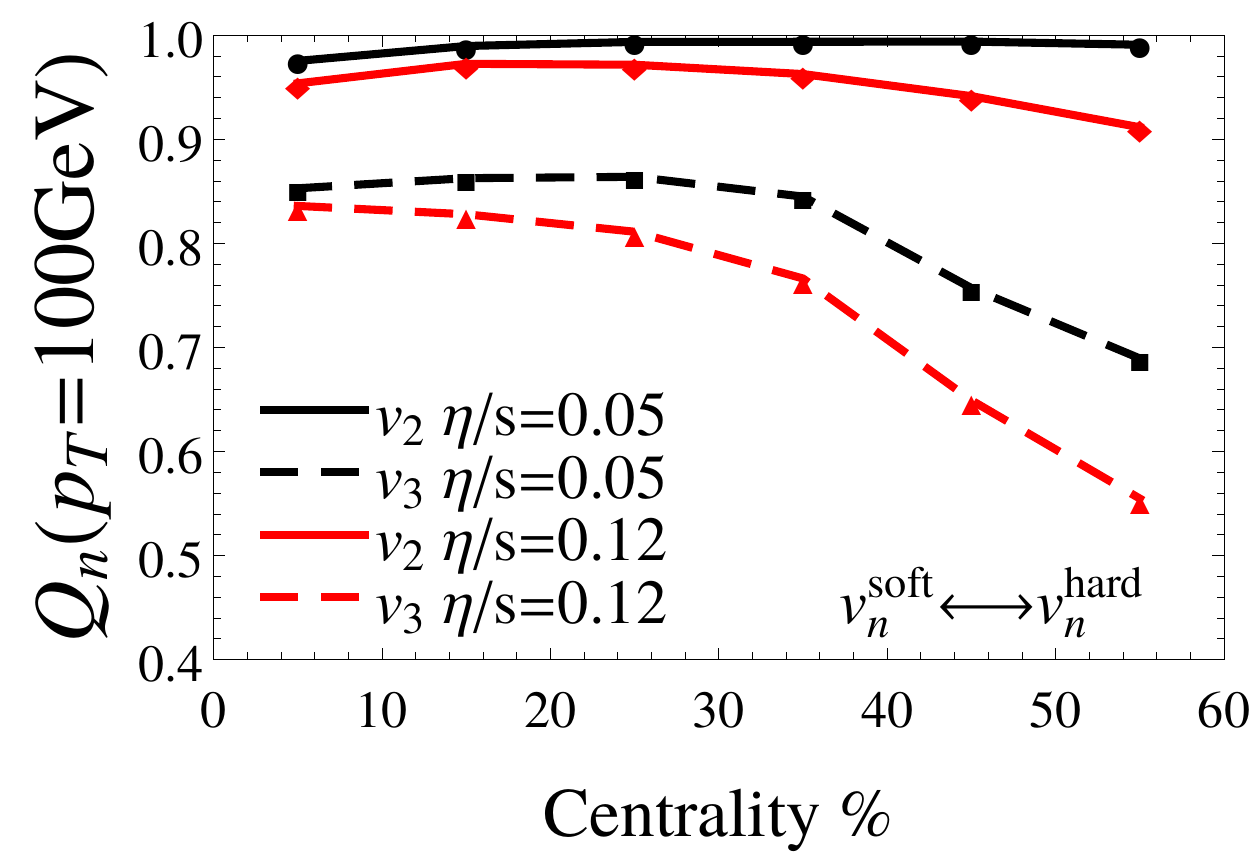}
\caption{(Color online) Pearson Coefficient, $Q_n$ between $v_n^{hard}$ and $v_n^{soft}$, across all centralities, described by Eq.\ (\ref{Qv2}) for $p_T=100$ GeV.}
\label{fig:quality100}
\end{figure}

\begin{figure}[ht]
\centering
\includegraphics[width=1\textwidth]{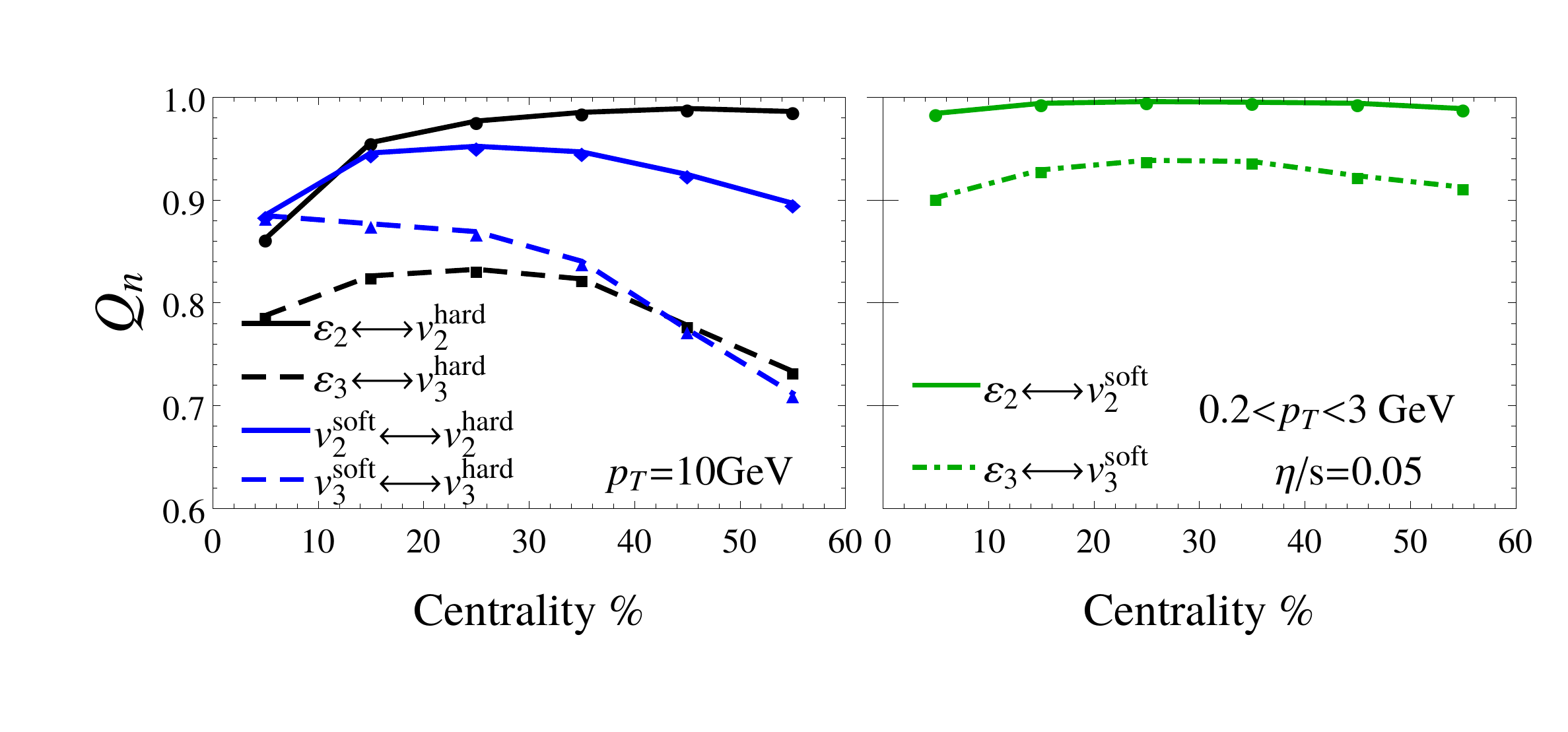}
\caption{(Color online) Pearson Coefficient, $Q_n$ between $v_n^{hard}$ and $\varepsilon_n$ for $p_T=100$ GeV (left) and  between $v_n^{soft}$ and $\varepsilon_n$ (right), across all centralities.}
\label{fig:qualityecc}
\end{figure}
Finally, we explore the correlation between initial eccentricities and soft flow harmonics.  While the soft flow harmonics are already known to be strongly correlated with the initial eccentricities, the linear response is not perfect. Thus, it is not straightforward to see if the eccentricities play a larger role in the formation of $v_n^{hard}$ or if $v_n^{soft}$ is more strongly correlated with $v_n^{hard}$. In Fig.\ \ref{fig:qualityecc} (left plot) we compare the Pearson coefficients between $v_n^{hard}$ and $\varepsilon_n$ to the coefficients found using $v_n^{hard}$ and $v_n^{soft}$. As a comparison we also show the very strong correlation between $\varepsilon_n$ and $v_n^{soft}$ on the right in Fig.\ \ref{fig:qualityecc}.  

As expected in the soft physics regime $\varepsilon_2$ and $\varepsilon_3$ are very strongly correlated with the final $v_2^{soft}$ and $v_3^{soft}$, respectively, in Fig.\ \ref{fig:qualityecc} (right panel).  However, the behavior of the flow harmonics in the hard physics region is not so simple.  At high $p_T$ the elliptic flow is primarily correlated with the eccentricities, and to a lesser extent with the $v_2^{soft}$. However, triangular flow demonstrates the opposite behavior where $v_3^{soft}$ is a much stronger predictor of $v_3^{hard}$ than the initial eccentricities with the exception of peripheral collisions.  This is a rather surprising effect that will be explored in a future study.  The effects of the deviation from perfect linear response have an interesting effect on the multiparticle cumulants, especially on the ratio of $v_2\{4\}/v_2\{2\}$, which will be detailed below.  In fact, we propose a new variable, $\Delta^{SH}_n$, in the next section that is only non-zero when there are deviations from perfect linear response.

%%%%%%%%%%%%%%%%%%%%%%%%%%%%%%%%%%%%%%%%%

\section{Predictions for $R_{AA}$ and harmonic flow cumulants at $\sqrt{s_{NN}}=5.02$ TeV}
\label{results}

\begin{figure*}[ht]
\centering
\includegraphics[width=1\textwidth]{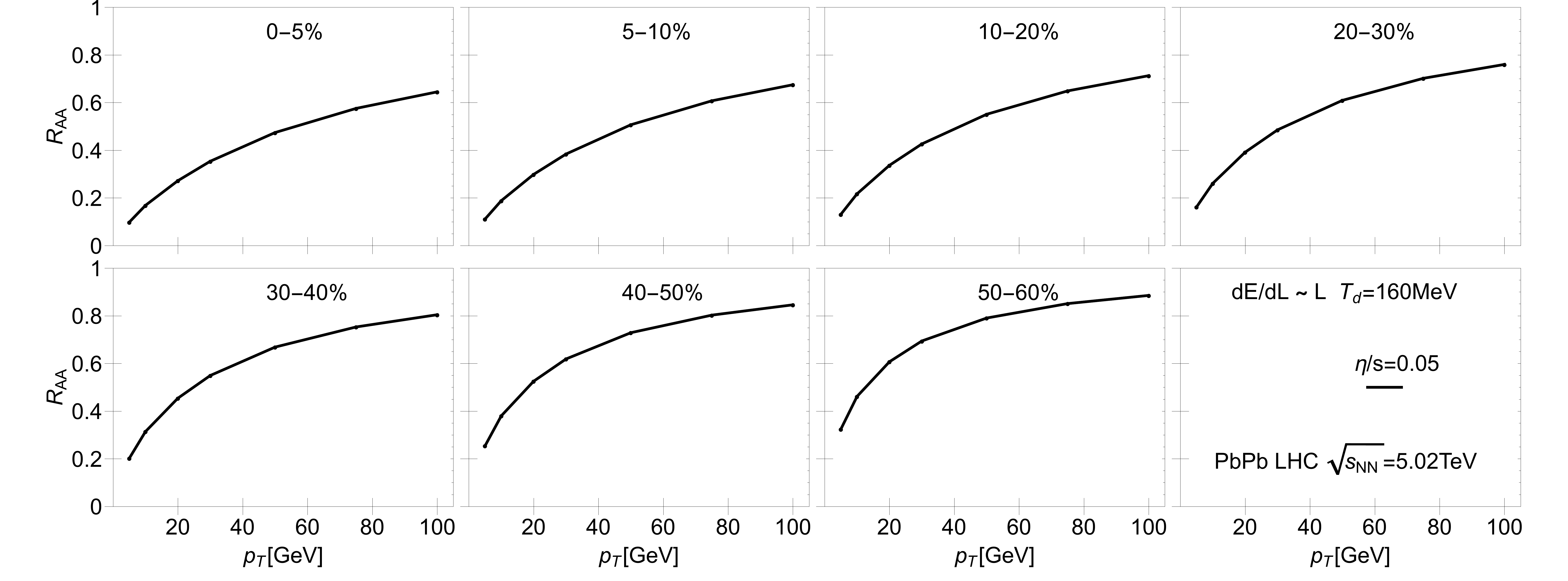} \\ 
\caption{(Color online) $R_{AA}(p_T)$ across centralities assuming linear path length dependence of the energy loss $dE/dL\propto L$, jet-medium decoupling temperature $T_d=160$ MeV, and  $\eta/s=0.05$ across all centralities and plotted up to $p_T=100$ GeV.  All values are calculated for PbPb LHC collisions at $\sqrt{s_{NN}}=5.02$ TeV.}
\label{fig:raa}
\end{figure*}

In Fig.\ \ref{fig:raa} we show our predictions for $\pi^0$ $R_{AA}(p_T)$ for our ``standard pQCD-like model" with a linear path length dependence $dE/dL\propto L$, jet-medium decoupling temperature $T_d=160$ MeV, and $\eta/s=0.05$ (the value that best describes the soft sector harmonic flow in our model, see Fig.\ \ref{fig:v2v3soft}). All errors in the plots presented in this  paper are statistical and they are calculated with jackknife resampling.  Across centralities there is very little change in the $p_T$ dependence of $R_{AA}$ though there is a modest increase around $p_T\sim 10$ GeV as one goes to more peripheral collisions. We checked the dependence of $R_{AA}(p_T)$ with $\eta/s$ and found that there was no visible difference between our standard choice of $\eta/s=0.05$ and the case where $\eta/s=0.12$ in Fig.\ \ref{fig:raa}, thus, $\eta/s=0.12$ is not shown here.
%%%%%%%%%%%%%%%%%%%%%%%%%%%%%%%%%%%%%%%%%%

%%%%%%%%%%%%%%%%%%%%%%%%%%%%%%%%%%%%%%%%%%

\begin{figure*}
\centering
\includegraphics[width=1\textwidth]{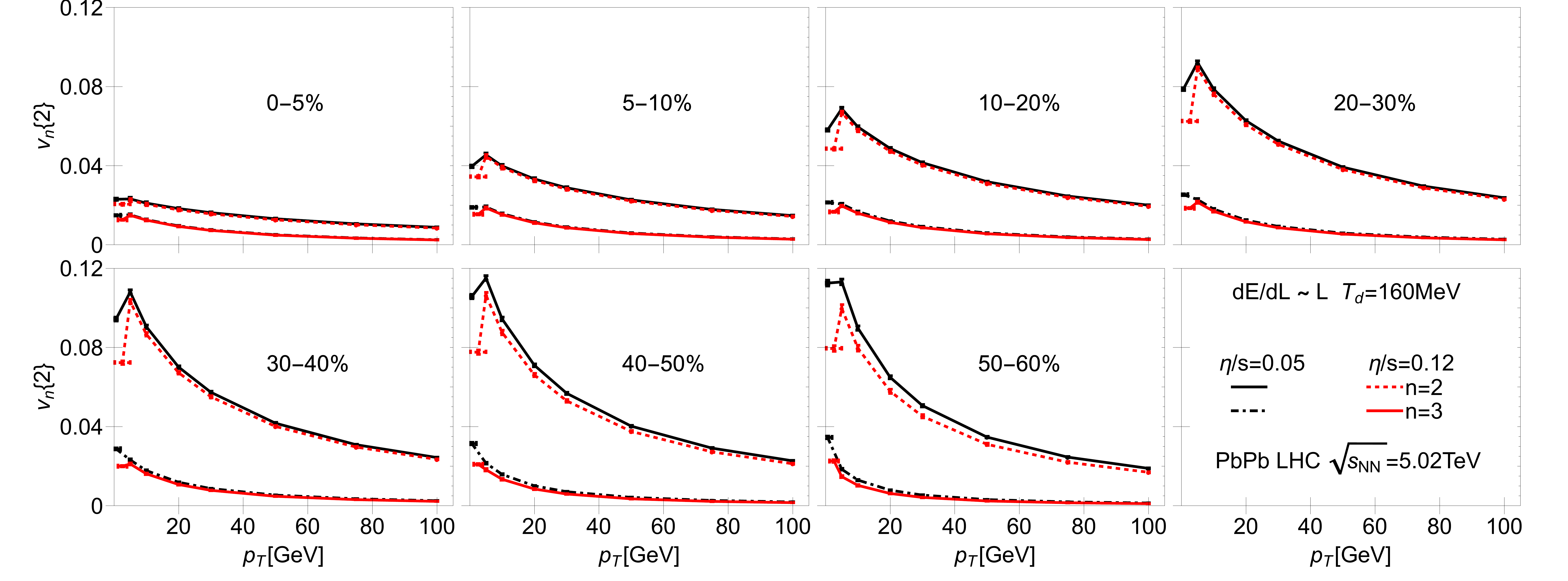}
\caption{(Color online) $v_2\{2\}(p_T)$ and $v_3\{2\}(p_T)$ across centralities computed assuming a linear path length dependence of the energy loss, $dE/dL\propto L$, and jet-medium decoupling temperature $T_d=160$ MeV. For the black curves $\eta/s=0.05$ while for the red curves $\eta/s=0.12$. All values are calculated for PbPb LHC collisions at $\sqrt{s_{NN}}=5.02$ TeV.}
\label{fig:v2v3}
\end{figure*}

In Fig.\ \ref{fig:v2v3} the high $p_T$ 2-particle cumulants of elliptic and triangular flow, $v_2\{2\}(p_T)$ and $v_3\{2\}(p_T)$, are shown across all centralities up to $p_T<100$ GeV. All the calculations of high $p_T$ cumulants in this paper are for $\pi^0$'s. Comparisons are shown between the results obtained with two different viscosities, $\eta/s=0.05$ and $\eta/s=0.12$, assuming a linear path length dependence for the energy loss $dE/dL\propto L$ and jet-medium decoupling temperature of $T_d=160$ MeV. The high $p_T$ flow harmonics show essentially no dependence on the viscosity for central to mid-central collisions and they appear to depend on only the initial eccentricities. For peripheral collisions, however, there is some viscosity dependence and, for the most peripheral $50-60\%$ collisions, it may even be possible to exclude one value of the viscosity via a comparison to data (depending on the size of the error bars) assuming that the initial eccentricity is known.  That being said, it is clear the viscosity effects in the soft sector, shown in Fig.\ \ref{fig:v2v3soft}, are at this time more appropriate to constrain the value of this transport coefficient using experimental data. However, for consistency, we expect that the high $p_T$ data would be more compatible with the lowest value of $\eta/s$ as well.  
%%%%%%%%%%%%%%%%%%%%%%%%%%%%%%%%%%%%%%%
%%%%%%%%%%%%%%%%%%%%%%%%%%%%%%%%%%%%%%%

From Figs.\ \ref{fig:raa}-\ref{fig:v2v3} it appears that $R_{AA}(p_T)$, $v_2\{2\}(p_T)$, and  $v_3\{2\}(p_T)$ at $p_T>10$ GeV have almost no sensitivity to the shear viscosity of the medium.  As shown in \cite{Noronha-Hostler:2016eow}, as well as in Section \ref{nonlinearity}, the eccentricities play the driving role among the bulk parameters in determining the high $p_T$ flow harmonics. In fact, it was shown in \cite{Noronha-Hostler:2016eow} that the high $p_T$ flow harmonics are sensitive to the choice of the initial conditions via its connection with the eccentricities. For instance, one could see in \cite{Noronha-Hostler:2016eow} that the more eccentric MCKLN initial conditions give larger $v_2\{2\}$ at high $p_T$ in comparison to the results found using MCGlauber. 

\begin{figure*}[ht]
\centering
\includegraphics[width=0.7\textwidth]{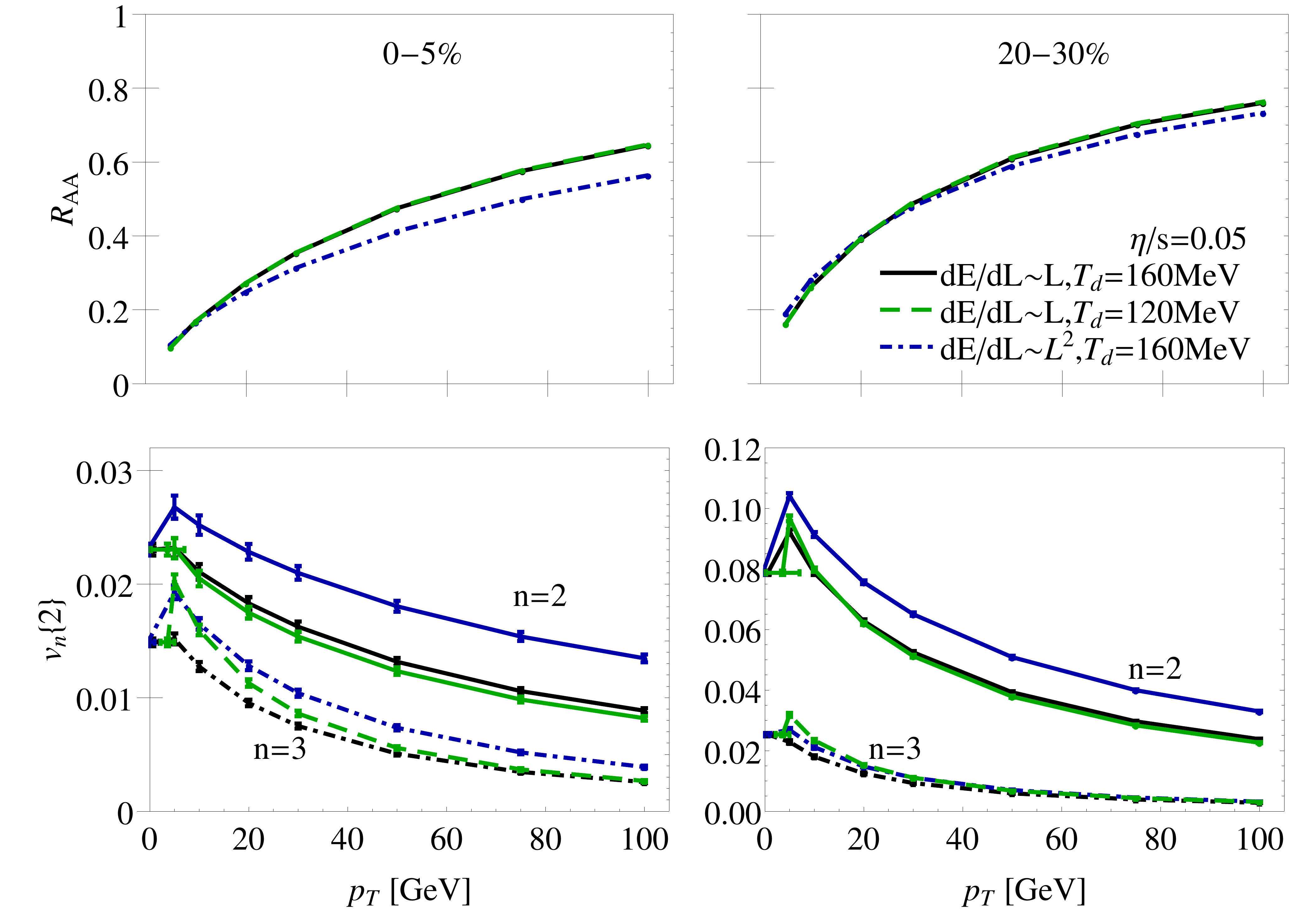}
\caption{(Color online) Variation of $R_{AA}(p_T)$, $v_2\{2\}(p_T)$, and $v_3\{2\}(p_T)$ with the path length dependence $dE/dL\propto L$ vs. $dE/dL\propto L^2$ and the jet-medium decoupling temperature $T_d=160$ MeV vs. $T_d=120$ MeV, keeping $\eta/s=0.05$. Only $0-5\%$ and $20-30\%$ centralities are shown. All values are calculated for PbPb LHC collisions at $\sqrt{s_{NN}}=5.02$ TeV.}
\label{fig:v2v3comp}
\end{figure*}

In Fig.\ \ref{fig:v2v3comp} we hold the viscosity constant at $\eta/s=0.05$ and vary either the path length dependence, i.e., $dE/dL\propto L$ vs. $dE/dL\propto L^2$ or the jet-medium decoupling temperature $T_d=160$ MeV vs. $T_d=120$ MeV for the centralities $0-5\%$ and $20-30\%$. We find no dependence of $R_{AA}(p_T)$ on the jet-medium decoupling temperature.  However, there is a clear splitting between the different choices for the path length dependence of the energy loss, linear vs. quadratic. If the error bars in the future LHC run 2 data are small enough on the $0-5\%$ centrality class it may be possible to exclude one of the possible path length dependences of the energy loss. 

For the flow harmonics there a modest increase in $v_3\{2\}(p_T)$ for the lower decoupling temperature while $v_2\{2\}(p_T)$ actually decreases slightly for $0-5\%$. Thus, the ratio of $v_2\{2\}(p_T)/v_3\{2\}(p_T)$ is sensitive to the value of $T_d$ though it remains to be seen if that effect is large enough to be constrained by experimental data. While the decoupling temperature has only a modest effect, the path length dependence plays a large role.  Both for $0-5\%$ and $20-30\%$ a quadratic path length dependence leads to a significantly larger $v_2\{2\}(p_T)$ and also a larger $v_3\{2\}(p_T)$. Therefore, between $R_{AA}(p_T)$,  $v_2\{2\}(p_T)$, and $v_3\{2\}(p_T)$ we expect it to be possible to further constrain the path length dependence of energy loss using the new LHC run 2 data. 

%%%%%%%%%%%%%%%%%%%%%%%%%%%%%%%%%%%%%%%

\begin{figure*}[ht]
\centering
\includegraphics[width=1\textwidth]{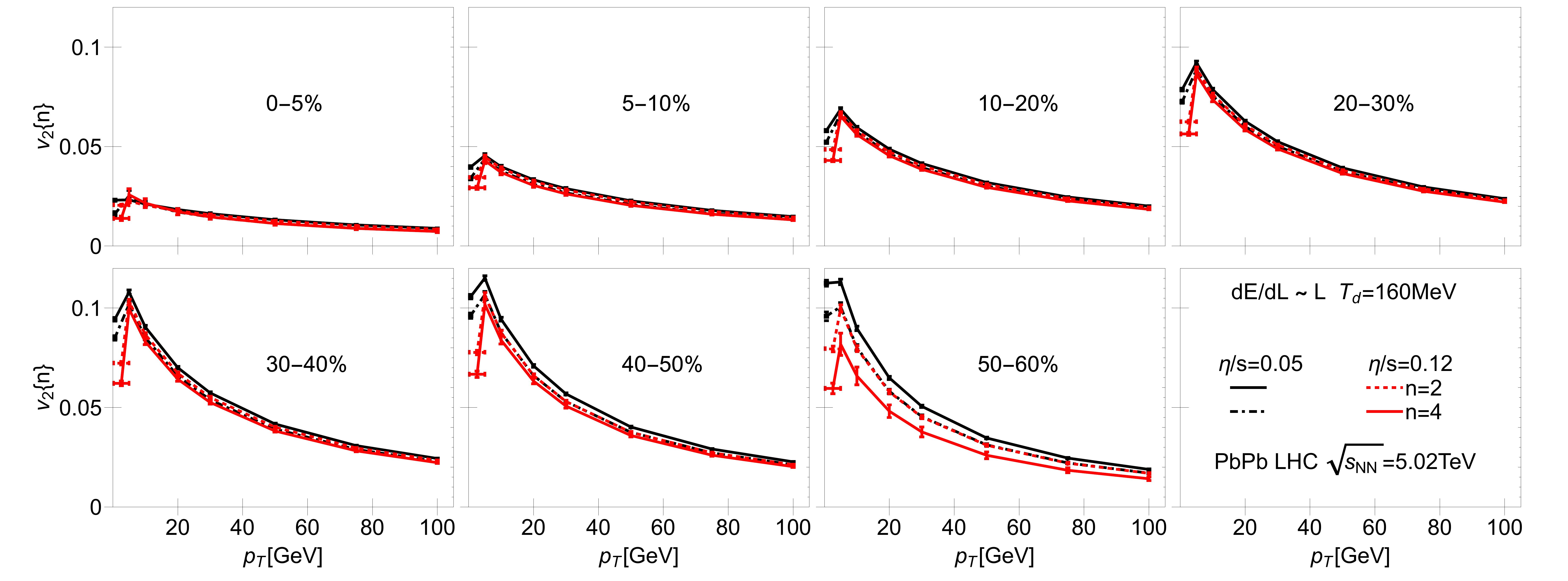}
\caption{(Color online) Cumulants $v_2\{2\}(p_T)$ and $v_2\{4\}(p_T)$ for $\eta/s=0.05$ and $\eta/s=0.12$ assuming a linear path length dependence $dE/dL\propto L$ and jet-medium decoupling temperature $T_d=160$ MeV across centralities. All values are calculated for PbPb LHC collisions at $\sqrt{s_{NN}}=5.02$ TeV.}
\label{fig:v2cum}
\end{figure*}

\begin{figure*}[ht]
\centering
\includegraphics[width=1\textwidth]{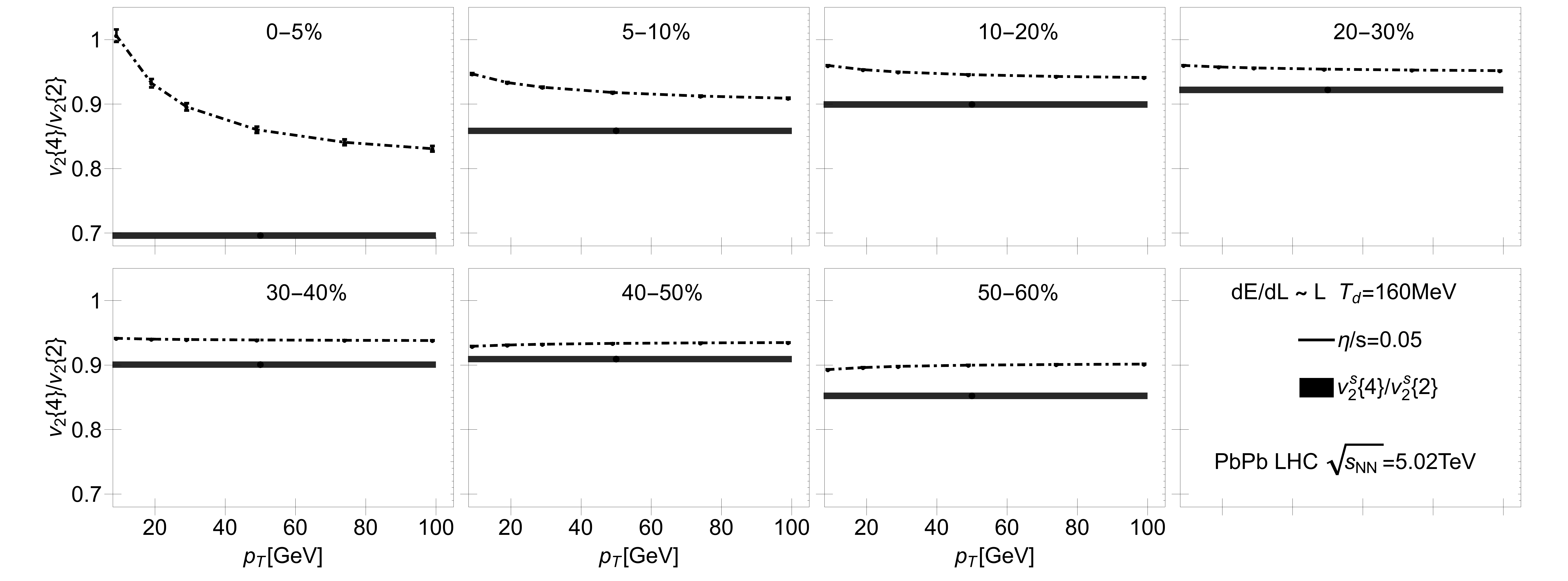}
\caption{(Color online) $v_2\{4\}(p_T)/v_2\{2\}(p_T)$ ratio across centralities for $\eta/s=0.05$, $dE/dL \propto L$, and jet-medium decoupling parameter $T_d=160$ MeV. The black band denotes the corresponding value of this ratio in the soft sector. All values are calculated for PbPb LHC collisions at $\sqrt{s_{NN}}=5.02$ TeV.}
\label{fig:v2v4ratlo}
\end{figure*}

\begin{figure*}[ht]
\centering
\includegraphics[width=1\textwidth]{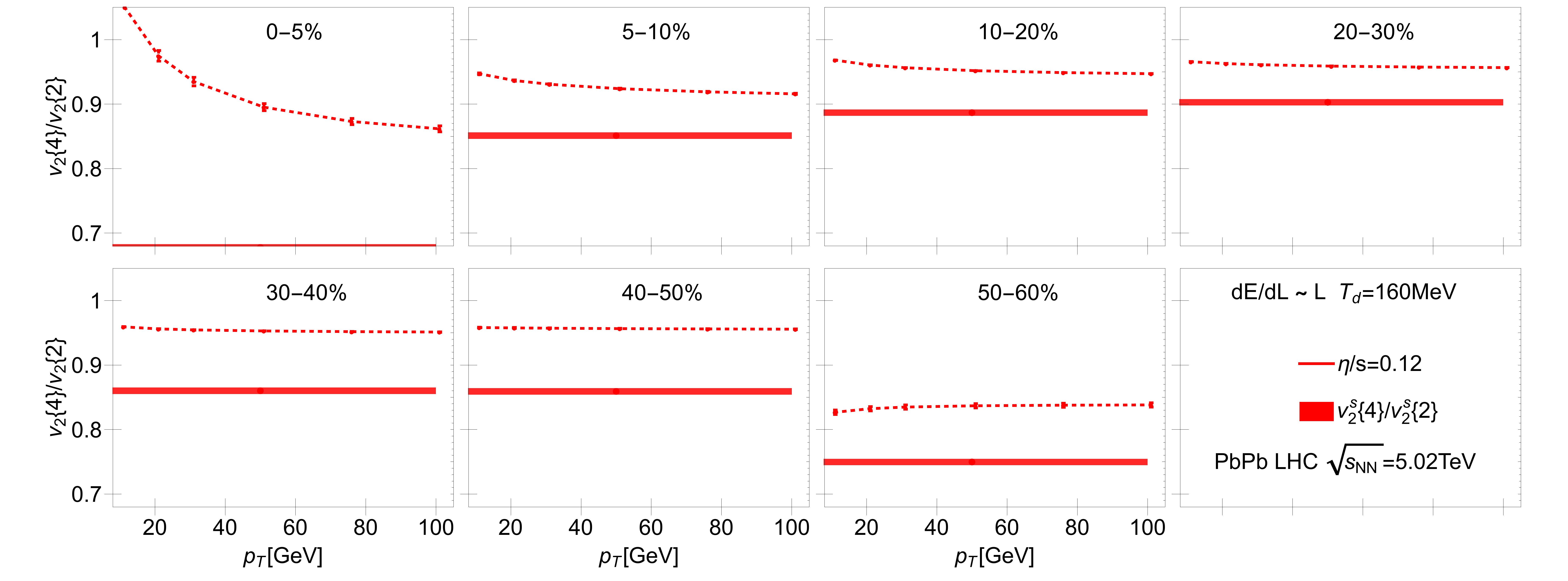}
\caption{(Color online) $v_2\{4\}(p_T)/v_2\{2\}(p_T)$ ratio across centralities for $\eta/s=0.12$, $dE/dL \propto L$, and jet-medium decoupling parameter $T_d=160$ MeV. The red band denotes the corresponding value of this ratio in the soft sector. All values are calculated for PbPb LHC collisions at $\sqrt{s_{NN}}=5.02$ TeV.}
\label{fig:v2v4rathi}
\end{figure*}

\begin{figure}[ht]
\centering
\includegraphics[width=0.45\textwidth]{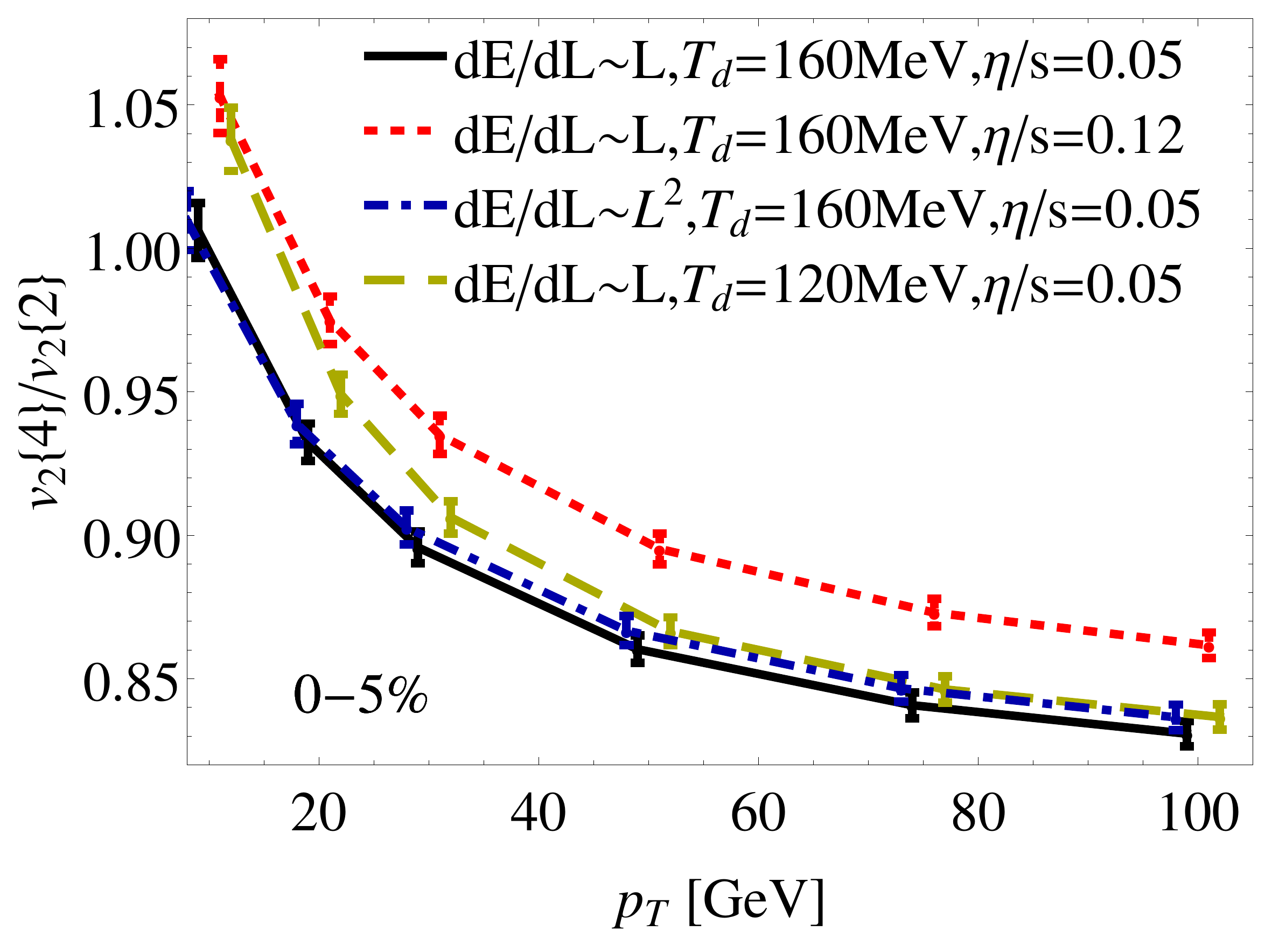}
\caption{(Color online) Variation in the $v_2\{4\}(p_T)/v_2\{2\}(p_T)$ ratio with changes in the path length dependence $dE/dL\propto L$ vs. $dE/dL\propto L^2$ as well as in the jet-medium decoupling temperature $T_d=160$ MeV vs. $T_d=120$ MeV, for $\eta/s=0.05$ and $\eta/s=0.12$. Only the $0-5\%$ centrality class is shown. All values are calculated for PbPb LHC collisions at $\sqrt{s_{NN}}=5.02$ TeV.}
\label{fig:v2v4ratcomp}
\end{figure}

In Fig.\ \ref{fig:v2cum} the results for $v_2\{2\}(p_T)$ and $v_2\{4\}(p_T)$ are shown for $\eta/s=0.05$ and $\eta/s=0.12$, assuming a linear path length dependence and $T_d=160$ MeV  across centralities.  As in Fig.\ \ref{fig:v2v3}, the effect of viscosity only appears in peripheral collisions.  Additionally, we find that the difference between $v_2\{2\}(p_T)$ and $v_2\{4\}(p_T)$  is smaller at high $p_T$ than at low $p_T$. In order to investigate this effect further the ratio of $v_2\{4\}(p_T)/v_2\{2\}(p_T)$ is shown in Figs.\ \ref{fig:v2v4ratlo}-\ref{fig:v2v4ratcomp}.  

%%%%%%%%%%%%%%%%%%%%%%%%%%%%%%%%%%%%%%%

In the low momentum region the ratio $v_2\{4\}/v_2\{2\}$ is often used to judge the strength of the fluctuations (large  $v_2\{4\}/v_2\{2\}$ indicates a narrower distribution whereas a smaller value indicates a wider distribution). Specifically, $v_2\{4\}/v_2\{2\}$ is related to the variance of $v_n^2$, $\sigma^2(v_n^2) \equiv \langle v_2^4 \rangle - \langle v_2^2 \rangle^2$, as
\begin{align}
\label{ratio}
\left(\frac{v_2\{4\}}{v_2\{2\}}\right)^4 &= 2 - \frac{\langle v_2^4 \rangle}{\langle v_2^2\rangle^2} \\
& = 1 - \frac {\sigma^2(v_n^2)} {\langle v_2^2 \rangle^2}.
\end{align}

The differential ratio $v_2\{4\}(p_T)/v_2\{2\}(p_T)$ involves a nontrivial correlation between $v_2$ at high and low $p_T$, and is therefore more complicated, as discussed in the next section.  However, if there is a perfect linear correlation between the integrated $v_2$ in each event and $v_2(p_T)$ at a fixed transverse momentum,  the ratios are equal.

Thus, in Fig.\ \ref{fig:v2v4ratlo} we plot the ratio across $p_T$ using our standard scenario with a linear path length dependence, $T_d=160$ MeV, and $\eta/s=0.05$.  One can see that there is a strong $p_T$ dependence in $v_2\{4\}(p_T)/v_2\{2\}(p_T)$ that approaches unity at $p_T\sim 10$ GeV for central collisions. As one goes towards more peripheral collisions this ratio becomes approximately constant with $p_T$. The black band is our corresponding prediction for the ratio $v_2\{4\}/v_2\{2\}$ in the soft sector, which is found to be smaller than the differential ratio $v_2\{4\}(p_T)/v_2\{2\}(p_T)$ at high $p_T$.  

As a comparison, in Fig.\ \ref{fig:v2v4rathi} we increase the viscosity to $\eta/s=0.12$ keeping the same path length dependence and decoupling temperature as in Fig.\ \ref{fig:v2v4ratlo} to see what effect it has on $v_2\{4\}(p_T)/v_2\{2\}(p_T)$.  One can see that the difference between the soft and hard ratios is more pronounced for the larger value of $\eta/s$ across all centralities. This is consistent with the results shown in Fig.\ \ref{fig:v2v3soft}.

Finally, in Fig.\ \ref{fig:v2v4ratcomp} a direct comparison is shown for different scenarios in the $0-5\%$ centrality bin, which has the largest deviation from the soft sector and the strongest $p_T$ dependence.  One can see that a larger viscosity gives the largest $v_2\{4\}(p_T)/v_2\{2\}(p_T)$ ratio and that this is the dominant effect.  Changing the path length dependence from $dE/dL\propto L$ to  $dE/dL\propto L^2$ has almost no effect on the ratio, which is interesting because this choice has a large effect on $R_{AA}(p_T)$, $v_2\{2\}(p_T)$, and $v_3\{2\}(p_T)$. In fact, looking at $v_2\{4\}(p_T)/v_2\{2\}(p_T)$ provides a method of checking the viscosity fit separately from the path length dependence. Finally, lowering the decoupling temperature to $T_d=120$ MeV gives a different dependence across $p_T$ for this ratio, which could not be clearly seen in previous plots.  Thus, in our model the $v_2\{4\}(p_T)/v_2\{2\}(p_T)$ ratio not only provides interesting information about the fluctuations at high $p_T$ but it also may be used to constrain the medium parameters. 

\subsection{Flow Fluctuations at high $p_T$}
The ratio of integrated cumulants $v_2\{4\}/v_2\{2\}$ is related to the variance of the $v_n^2$ distribution (see Eq.~\eqref{ratio}), and goes to unity as the fluctuations vanish and the variance goes to zero.  The differential cumulants $v_n\{k\}(p_T)$, on the other hand, represent a non-trivial correlation between $v_n$ at different transverse momenta (see Eqs.~\eqref{dn} and \eqref{cn}).   If there are no fluctuations at all (hard or soft), one again obtains a ratio $v_2\{4\}(p_T)/v_2\{2\}(p_T)\sim 1$.  However, the converse is not true --- a value of 1, such as that seen in central collisions at lower $p_T$,  does \textit{not} necessarily imply a lack of fluctuations in either the hard or soft sector --- and unlike the case for integrated cumulants, a value greater than 1 is possible.

%If one considers the paradigm of integrated flow harmonics in the soft sector, then one would assume that as $v_2\{4\}(p_T)/v_2\{2\}(p_T)\rightarrow 1$ the high $p_T$ flow harmonics cease to fluctuate and all converge roughly to the same value.  As it turns out, that would only be true if the soft sector \textit{also} was not fluctuating.  
In fact, flow fluctuations have been well-documented not only in soft physics,  but also already from experimental data that clear fluctuations in $v_2\{2\}$ have been measured up to $p_T~15$ GeV \cite{Aad:2015lwa}. 
In our model, we can see clear fluctuations in the scatter plot in Fig. \ref{fig:scat020} despite the fact that the ratio is $v_2\{4\}(p_T)/v_2\{2\}(p_T)\sim 1$  in Fig. \ref{fig:v2v4ratlo}.

%Let us assume, despite all evidence to the contrary, that there were no flow fluctuations at high $p_T$.  In that case we start with
%\begin{equation}
%v_2\{4\}(p_T)=\frac{2 \langle |v_n^{s}|^2\rangle  \langle v^{s}_n v_n^{h}(p_T)\cos\left[n\left(\psi^{s}_n-\psi^{h}_n(p_T)\right]\right) \rangle-\langle |v^{s}_n|^3 v_n^{h}(p_T)\cos\left[n\left(\psi^{s}_n-\psi^{h}_n(p_T)\right]\right) \rangle}{(v_n^{s}\{4\})^{3/4}} \nonumber
%\end{equation}
%where we ignore multiplicity weighing for a moment.
%If there are no fluctuations at high $p_T$, we can assume there is only one value for $v_n^h(p_T)=\langle v_n^h(p_T)\rangle$,  then
%\begin{eqnarray}\label{eqn:stillfluc}
%v_n\{4\}(p_T)&\rightarrow& v_n^h(p_T)\frac{2 \langle |v_n^{s}|^2\rangle  \langle v^{s}_n  \rangle-\langle |v^{s}_n|^3  \rangle}{(2 \langle |v_n^{s}|^2\rangle ^2-\langle |v^{s}_n|^4  \rangle)^{3/4}} 
%\end{eqnarray}
%However, one would never arrive at $v_n\{2\}(p_T)\sim v_n\{4\}(p_T)$ because the soft sector is still fluctuating, which will always affect the right-hand side of Eq.\ (\ref{eqn:stillfluc}).  

A clearer way to study the difference between harmonic flow fluctuations at low and high $p_T$ may be obtained using the observable
\begin{eqnarray}
\Delta^{SH}_n(p_T)&\equiv & \underbrace{\frac{\langle  v_n^4\rangle}{\langle  v_n^2\rangle^2}}_{\text{soft fluctuations}}- \underbrace{\frac{\langle v_n^2  V_n V_n^*(p_T)\rangle}{\langle v_n^2\rangle\langle  V_n V_n^*(p_T)\rangle}}_{\text{hard fluctuations}}\nonumber\\
&=& \underbrace{\left(\frac{v_n\{2\}}{v_n\{4\}}\right)^5\left[ \frac{v_n\{4\}(p_T) }{v_n\{2\}(p_T) }-\frac{v_n\{4\} }{v_n\{2\} }\right]}_{\text{Experimental observable}}\label{eqn:deltash}
\end{eqnarray}
where the relationship above is exact (before any centrality rebinning), as shown in Appendix \ref{appen}.
If the fluctuations of high $p_T$ elliptic flow were exactly given by the soft fluctuations in a linearly correlated manner (on an event by event basis), i.e., $V_n(p_T) \to \chi_n(p_T)V_n$ with $\chi_n(p_T)$ being the same for all events in the given centrality class, then $\Delta^{SH}_n(p_T)$ would be identically zero for all $p_T$. The discussion in Section \ref{nonlinearity} shows that this should not be the case and, in fact, one can already see in Figs.\ \ref{fig:v2v4ratlo} and \ref{fig:v2v4rathi} that such quantity is nonzero. This implies that Eq.\ (5) of \cite{Noronha-Hostler:2016eow}, which was derived assuming linear response, cannot be used to obtain the correct magnitude of the effects of event-by-event fluctuations on $v_2\{2\}(p_T)$. 

\begin{figure}[ht]
\centering
\includegraphics[width=0.45\textwidth]{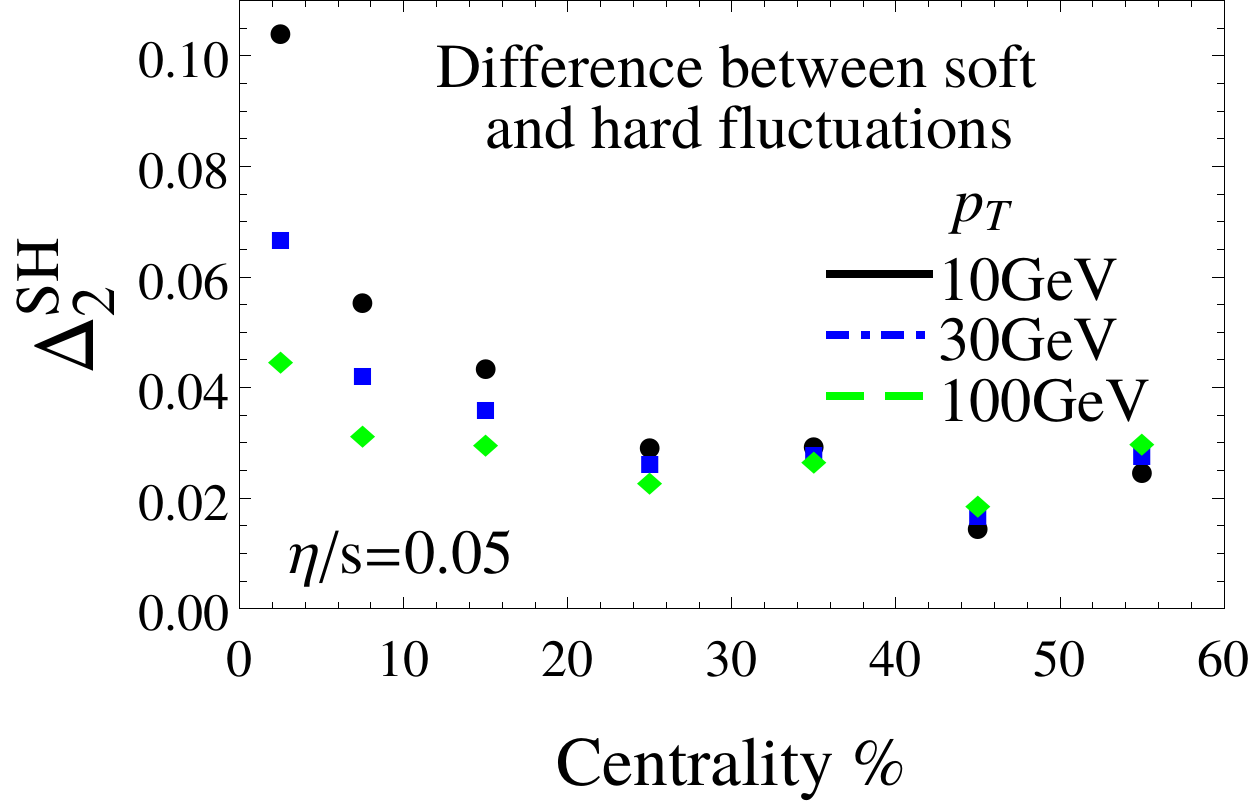}
\caption{(Color online) Difference between the soft and hard fluctuations, $\Delta^{SH}_2$, defined in Eq.\ (\ref{eqn:deltash}) for $\eta/s=0.05$ across all centralities and for three values of $p_T$. }
\label{fig:deltasheta05}
\end{figure}

\begin{figure}[ht]
\centering
\includegraphics[width=0.45\textwidth]{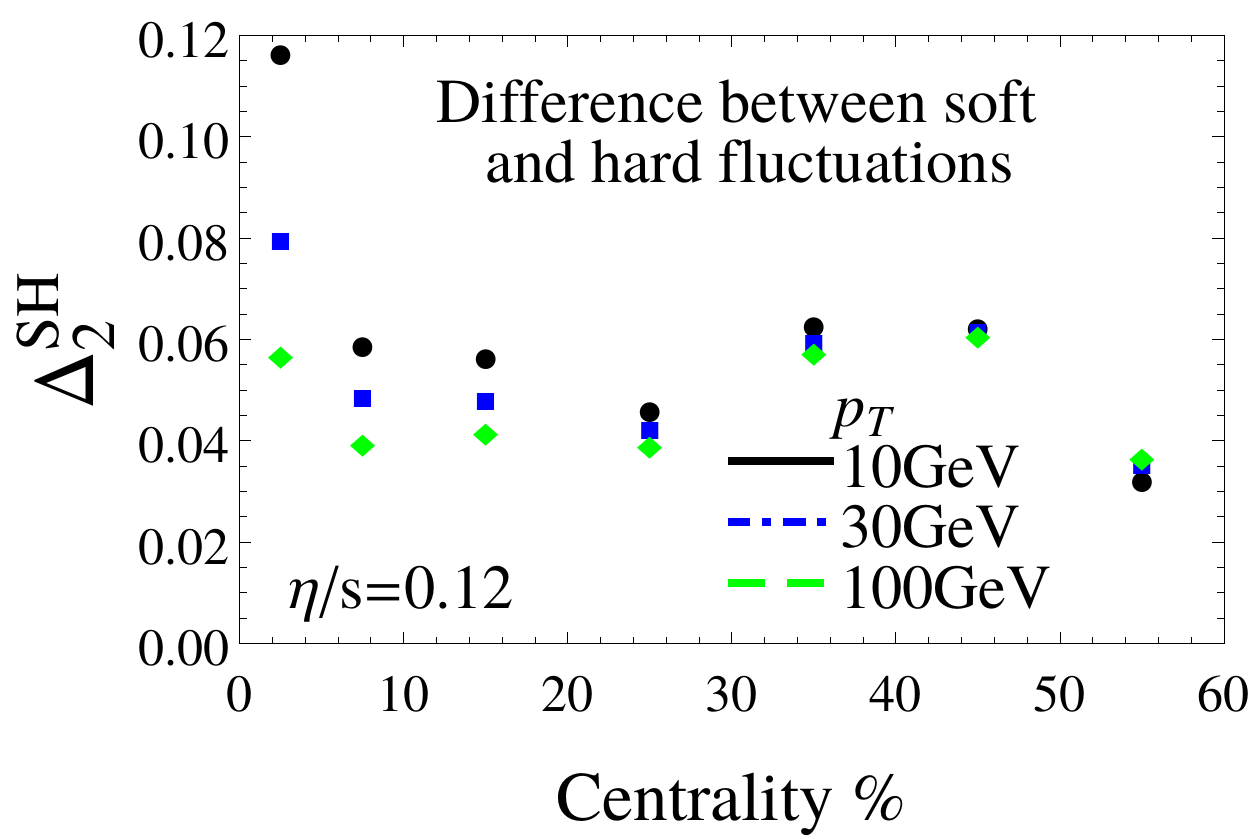}
\caption{(Color online) Difference between the soft and hard fluctuations, $\Delta^{SH}_2$, defined in Eq.\ (\ref{eqn:deltash}) for $\eta/s=0.12$ across all centralities and for three values of $p_T$. }
\label{fig:deltasheta12}
\end{figure}

In the soft sector, the decorrelation of $v_n$ at different $p_T$ is studied with 2-particle correlations via the factorization breaking ratio $r_n$ \cite{Gardim:2012im}, or via principle component analysis (PCA) \cite{Bhalerao:2014mua}.  However, these analyses require a measurement of a two-particle correlation with both particles at a fixed $p_T$.   For transverse momenta above 10 GeV, this is unfeasible, and it is therefore necessary to study correlations where only 1 of the particles is restricted to a high $p_T$ bin, as we propose here.  

Figs.\ \ref{fig:deltasheta05}-\ref{fig:deltasheta12} show that $\Delta^{SH}_2(p_T)$ possess a clear dependence on the centrality class and the value of $p_T$.  For the most central collisions and $p_T=10$ GeV we find the maximum difference between the fluctuations in the soft and hard sectors.  As one increases $p_T$ the fluctuations in the soft and hard sectors are more similar (and that is relatively constant across centrality). However, we note that even at very high $p_T$ the assumption of a linear relationship between the high and low $p_T$ elliptic flows does not hold since $\Delta^{SH}_2 \neq 0$. Furthermore, a comparison between Figs.\ \ref{fig:deltasheta05} and \ref{fig:deltasheta12} shows that this quantity is also sensitive to the viscosity of the medium. In fact, the difference between the fluctuations in the hard and soft sectors are found to increase with $\eta/s$, which is expected given the large sensitivity of the soft flow cumulants with viscosity (see Fig.\ \ref{fig:v2v3soft}).

\section{Soft-hard event plane correlation}
\label{decorrelation}

In Section \ref{nonlinearity} the linear relationship between the soft and the hard harmonic flow coefficients was explored in terms of the magnitude of the flow vectors (in the scatter plots of Figs.\ \ref{fig:scat020}-\ref{fig:scat4060}) and the entire flow vector through linear correlation coefficients $Q_n$, Eq.\ \eqref{Qv2}. 
One can also visualize how their event plane angle changes at high $p_T$.  As one goes out to higher and higher $p_T$ it is nontrivial to assume that the event plane angle of the integrated soft flow harmonic is correlated with the corresponding quantity at $p_T=100$ GeV. The correlation function between soft and hard flow harmonics in Eqs.\ (\ref{dnj2}) and (\ref{dnj4}) necessarily contains a cosine term of the difference between their event plane angles. Thus, any degree of decorrelation between these angles decreases the harmonic flow cumulants. In \cite{Jia:2012ez} it was suggested that this decorrelation effect is extremely small for $v_2\{2\}$ whereas $v_3\{2\}$ should be more strongly affected by the decorrelation of the corresponding event plane angles.  

\begin{figure}[ht]
\centering
\includegraphics[width=0.45\textwidth]{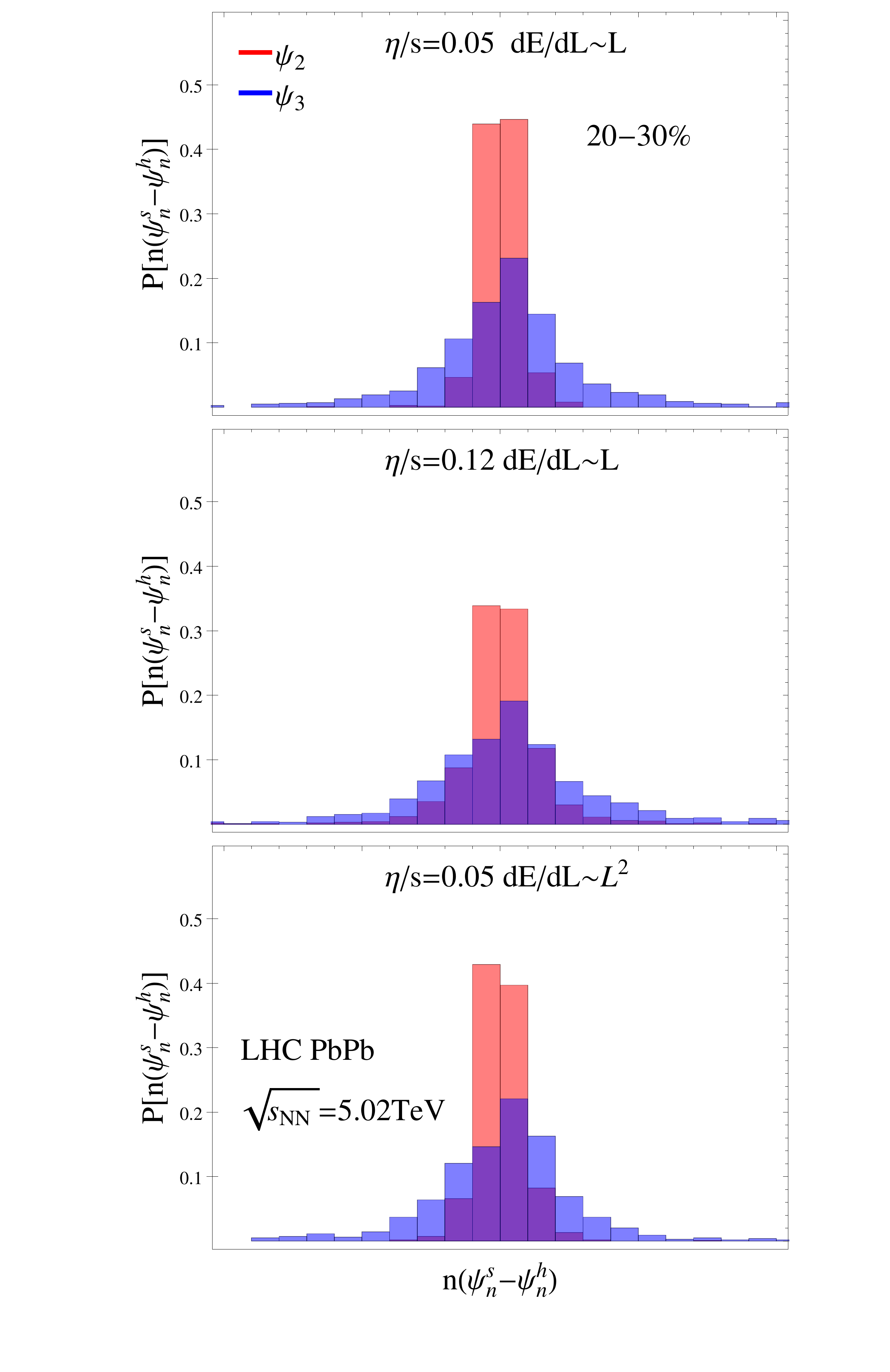}
\caption{(Color online) Distribution of the difference in the event plane angles at low and high $p_T$, $P(n[\psi_n^{soft}-\psi_n^{hard}(p_T)])$, at $20-30\%$ centrality for LHC PbPb $\sqrt{s_{NN}}=5.02$ TeV. }
\label{fig:EPdis}
\end{figure}

Because in this study we have an order of magnitude larger statistics as well as a wider range in parameter variation than \cite{Noronha-Hostler:2016eow}, we can determine both how large of an effect the event plane decorrelation has on the flow harmonics and what aspects of the medium influence this decorrelation.  In Fig.\ \ref{fig:EPdis}, the difference in the event plane angles at low and high $p_T$, $P(n[\psi_n^{soft}-\psi_n^{hard}(p_T)])$, at $20-30\%$ centrality is shown for $n=2$ and $n=3$.  One can clearly see that there is a very strong correlation between the soft and hard angles for elliptic flow whereas for triangular flow the angles are less correlated, which suppresses $v_3\{2\}(p_T)$.  
It is also interesting to note that the path length dependence of the energy loss has essentially no influence on this result whereas a larger shear viscosity leads to a larger decorrelation in the event plane angles between the soft and the hard sectors.  

\begin{figure}[ht]
\centering
\includegraphics[width=0.45\textwidth]{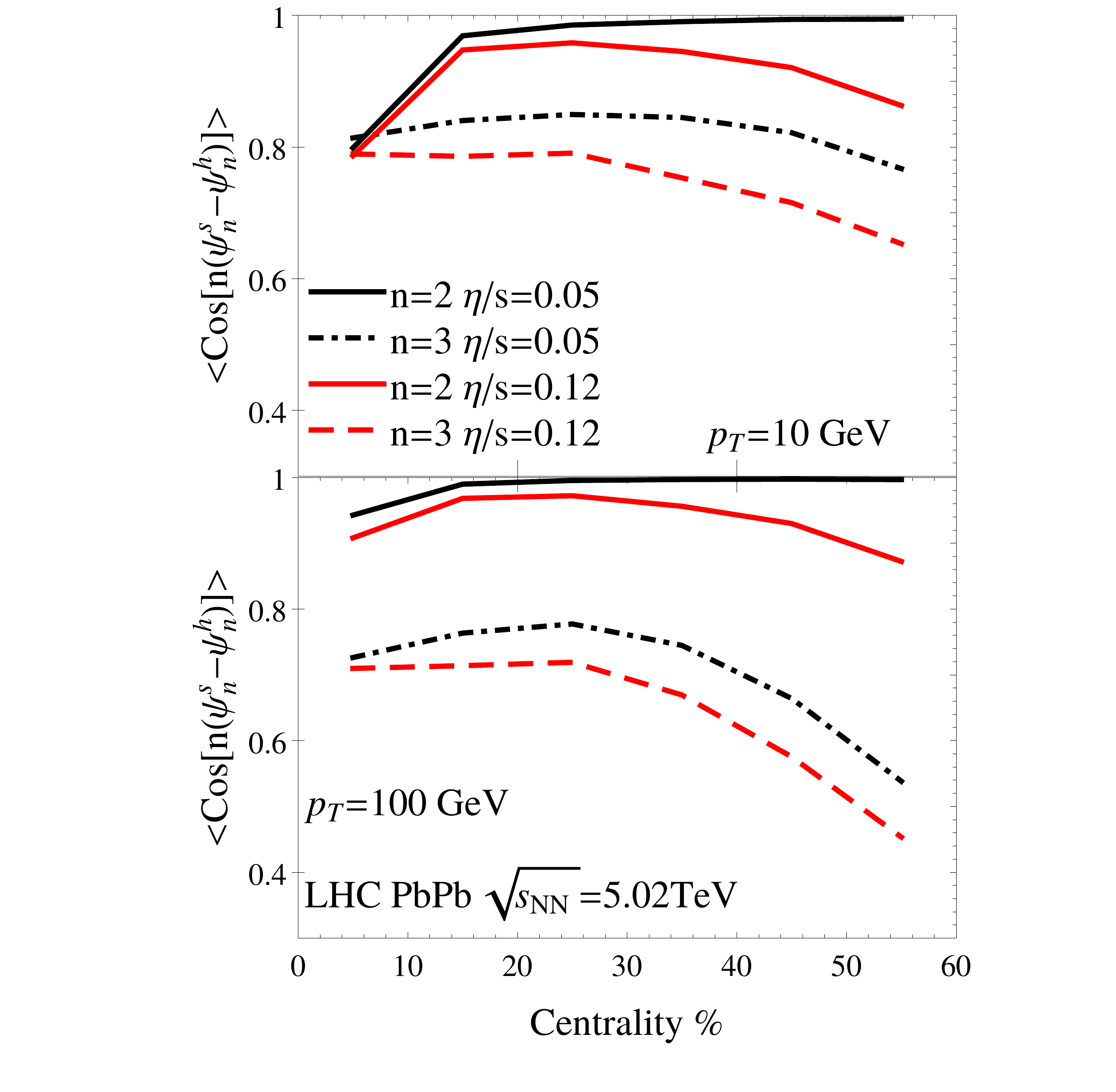}
\caption{(Color online) $\langle \cos n\left[\psi_n^{soft}-\psi_n^{hard}(p_T)  \right]\rangle$, for $n=2$ and $n=3$, across all centralities for two different viscosities $\eta/s=0.05$ and $\eta/s=0.12$ for LHC PbPb collisions at $\sqrt{s_{NN}}=5.02$ TeV. }
\label{fig:meanEP}
\end{figure}

To see this more clearly we plot in Fig.\ \ref{fig:meanEP} the mean of the cosine term in the correlation function across centralities, $\langle \cos n\left[\psi_n^{soft}-\psi_n^{hard}(p_T)  \right]\rangle$.  From Fig.\ \ref{fig:meanEP} one can conclude that the decorrelation of the event plane angles is strongly affected by viscosity. A larger shear viscosity suppresses $\langle \cos n\left[\psi_n^{soft}-\psi_n^{hard}(p_T)  \right]\rangle$ the most in central and peripheral collisions.  The event plane of triangular flow is especially sensitive to this effect.  A variation of the path length dependence of the energy loss did not change this result.  Thus, our results not only confirm \cite{Jia:2012ez} but we also find that event plane angle decorrelation at high $p_T$ may be used as a probe of the properties of the medium given its strong dependence with viscosity.  

%%%%%%%%%%%%%%%%%%%%%%%%%%%%%%%%%%%%%%%

%%%%%%%%%%%%%%%%%%%%%%%%%%%%%%%%%%%%%%%

\section{Conclusions and Outlook}
\label{final}

In this paper we used a combination of event-by-event relativistic hydrodynamics, given by the v-USPhydro code, with an energy loss model, BBMG, to make predictions for the high $p_T$ dependence of $R_{AA}(p_T)$, $v_2\{2\}(p_T)$, $v_3\{2\}(p_T)$, and $v_2\{4\}(p_T)$ of neutral pions in PbPb collisions at $\sqrt{s_{NN}}=5.02$ TeV, which can be tested against the upcoming results from run 2 data at LHC. Aside from $R_{AA}(p_T)$, none of the 2- and 4-particle cumulants discussed in this paper can be computed without considering the effects of event-by-event viscous hydrodynamics in jet energy loss calculations. In fact, as discussed in Section \ref{bigsection}, in meaningful comparisons to experimental data, the inclusion of event-by-event fluctuations in the theoretical calculation of high $p_T$ flow harmonics is not an option - it is rather mandatory given our current understanding of the bulk evolution of the QGP and the very definition of these observables via event-by-event correlations between soft and hard hadrons. The same reasoning applies to the heavy quark sector (see, e.g., \cite{Christiansen:2016uaq,Prado:2016xbq} for ideas on how to combine event shape engineering with light and heavy flavor and \cite{Cao:2013ita} for the influence of the initial state on heavy flavor flow) and results in this direction will be presented soon. We note that in \cite{Nahrgang:2014vza,Nahrgang:2016lst} heavy flavor triangular flow was calculated using the event plane method with an event-by-event ideal hydrodynamic background. In this regard, it would be interesting to study flow harmonic cumulants in the heavy flavor sector following the study done here using event-by-event viscous hydrodynamics.

In order to investigate how our results vary with the assumptions regarding the BBMG energy loss model, we varied the path length dependence of the energy loss, $dE/dL \propto L$ to $dE/dL \propto L^2$. We found a sensitivity of $R_{AA}(p_T)$, $v_2\{2\}(p_T)$ and $v_3\{2\}(p_T)$ with this change. From the combination of the three experimental observables it may be possible to constrain the type of path length dependence of the energy loss using LHC run 2 data.  The highest sensitivity occurs for the most central $<5\%$ collisions where from  Fig.9   
resolving $L^1$ from $L^2$ energy loss will require reduction of systematic errors on both 
$R_{AA}$ and $v_2$ to below 0.005.   Furthermore, we also tested two different shear viscosities $\eta/s=0.05$ and $\eta/s=0.12$, whose variation has no visible effect on $R_{AA}(p_T)$, though we found that $v_2\{2\}(p_T)$ sees a small suppression for the more peripheral collisions. Overall, the high $p_T$ flow harmonics are found to be much less sensitive to variations of the viscosity than their soft counterparts.

The last parameter that we varied in our study was the jet-medium decoupling temperature, $T_d$. This phenomenological scale sets the minimum temperature in the hadronic phase below which energy loss is taken to be zero. In this paper we varied $T_d$ between $120$ MeV and $160$ MeV and found that most experimental observables appear to be relatively insensitive to $T_d$, with the exception of the $v_2\{2\}(p_T)$ to $v_3\{2\}(p_T)$ relationship. Triangular flow requires longer system times to build up and, therefore, if the jet is coupled to the medium for a longer period of time then this will enhance $v_3\{2\}(p_T)$.

Here we also investigated the correlation between the soft and the hard event plane angles and its connection to viscosity. We confirm previous results found in \cite{Jia:2012ez} and \cite{Noronha-Hostler:2016eow} that the elliptic flow event plane angle at high $p_T$ is strongly correlated to the soft elliptic flow event plane angle. Moreover, we find that an increase in viscosity decorrelates the soft and the hard event plane angles, which is an interesting effect to explore in the future. 

Our model allows for the calculation of the difference between the harmonic flow fluctuations in the hard and in the soft sectors, as discussed in detail in Section \ref{nonlinearity}. We found that the linear correlation between high $p_T$ elliptic flow and the initial $\varepsilon_2$ is weaker than the correlation between $v_2^{soft}$ and $\varepsilon_2$. Also, triangular flow scales better with $v_3^{soft}$ than with the actual eccentricity $\varepsilon_3$. This deviation from perfect linear scaling of high $p_T$ elliptic flow affects the ratio $v_2\{4\}(p_T)/v_2\{2\}(p_T)$ and experimental verification of this effect can be done through the measurement of the quantity $\Delta_{SH}(p_T)$ defined in \eqref{eqn:deltash}, which involves the difference between the soft and hard fluctuations. This quantity depends on the initial conditions and the viscosity of the medium (differently than its soft counterpart) though it does not display a strong sensitivity to the choice for the path length dependence of the energy loss.

In the first attempt of combining event-by-event hydrodynamics with jets \cite{Noronha-Hostler:2016eow}, the initial conditions were varied and shown to play a significant role in the description of $v_2\{2\}(p_T)$. MCGlauber initial conditions, which have a smaller $\varepsilon_2\{2\}$ than that found in MCKLN initial conditions, consistently were at the low end of the $v_2\{2\}(p_T)$ error bars for LHC Run 1. However, MCKLN initial conditions were found to give a reasonable description of the experimental data at $\sqrt{s_{NN}}=2.76$ TeV. Due to this result, only MCKLN initial conditions were explored in this study. However, the choice of the initial condition for the hydrodynamic evolution plays a nontrivial role in the study of high $p_T$ flow harmonics and this subject certainly deserves further investigation. Because viscous effects here are small, the initial conditions have a dominant effect at high $p_T$ for $v_2\{2\}$ and $v_3\{2\}$. Thus, high $p_T$ flow harmonic provides a novel (and independent) opportunity to constrain the initial conditions. We hope to see future analyses using Bayesian techniques \cite{Bernhard:2016tnd} in viscous hydrodynamics + jet models to determine, in a systematic manner, the allowed range of model parameters that simultaneously describe soft and hard flow harmonics.

Finally, in this paper we went through all the details needed to perform this novel type of theoretical calculations of high $p_T$ flow harmonics event-by-event, including the definition of harmonic flow cumulants at high $p_T$, which considerably extends the initial study performed in \cite{Noronha-Hostler:2016eow}. Using this knowledge, a similar study could be carried out for hard sector observables using other types of initial conditions, bulk hydrodynamic evolution models (going, for instance, from 2+1 to full 3+1 hydrodynamics) and more realistic energy loss models such as \cite{Gyulassy:2000er,Xu:2014ica,Xu:2015bbz,Guo:2000nz,Wang:2001ifa,Majumder:2009ge,Qin:2007rn,Schenke:2009gb,Zapp:2012ak}. The combination of event-by-event viscous hydrodynamics and jet quenching models is indispensable for calculating triangular flow and multiparticle cumulants of flow harmonics at high $p_T$. This provides a novel tool that can be used to understand the correlation between the hard and the soft sectors of heavy ion collisions giving, thus, valuable insight onto how jets interact with the quark-gluon plasma.

%%%%%%%%%%%%%%%%%%%%%%%%%%%%%%%%%%%%%%%%%%

 \section*{Acknowledgements}

%%%%%%%%%%%%%%%%%%%%%%%%%%%%%%%%%%%%%%%%

% FORA TEMER GOLPISTA!

%%%%%%%%%%%%%%%%%%%%%%%%%%%%%%%%%%%%%%%%

The authors would like to thank W.~Li, Q.~Wang, M.~Guilbaud, A.~Timmins, A.~Suaide, C.~Prado, S.~Mohapatra, and Y.~Zhou  for discussions on how to compare theoretical calculations of high $p_T$ harmonic flow to experimental data. We thank X.-N.~Wang for providing the LO pQCD parton cross sections. J.N.H. was supported by the National Science Foundation under grant
no. PHY-1513864. J.N.H. and B.B. acknowledge the use of the Maxwell
Cluster and the advanced support from the Center of Advanced
Computing and Data Systems at the University
of Houston to carry out the research presented here. J.N. thanks the University 
of Houston for its hospitality and Funda\c c\~ao de Amparo \`a Pesquisa do Estado de 
S\~ao Paulo (FAPESP) and Conselho Nacional de Desenvolvimento Cient\'ifico e Tecnol\'ogico (CNPq) for support. 

%%%%%%%%%%%%%%%%%%%%%%%%%%%%%%%%%%%%%%%%

\appendix
\section{Derivation of $\Delta^{SH}_n$}\label{appen}

In the following, we will use capital $V_n$ to indicate the vector form of the flow  harmonics and the magnitude of a flow harmonic will be written as $v_n$.  In the language of soft vs. hard physics, it is understood that a cumulant $v_n\{m\}$ is a soft flow harmonic where as $v_n\{m\}(p_T)$ is the flow harmonic cumulant for the hard sector.  

In order to understand the high $p_T$ fluctuations further, we rewrite the ratio  $v_2\{4\}(p_T)/v_2\{2\}(p_T)$  using Eqs. (\ref{eqn:cumusimp2}-\ref{eqn:cumusimp4}) in the simplified vector form (defined in Section \ref{bigsection}) such that:
\begin{eqnarray}
\frac{v_n\{4\}(p_T) }{v_n\{2\}(p_T) }&=&\frac{2\langle  V_n V_n^*(p_T)\rangle \langle v_n^2\rangle -\langle v_n^2  V_n V_n^*(p_T)\rangle }{v_n\{4\}^3 }\frac{v_n\{2\} }{\langle V_n  V_n^*(p_T)\rangle}\nonumber\\
&=&\frac{v_n\{4\} }{v_n\{2\} } \left[\frac{2v_n\{2\}^4}{v_n\{4\}^4 }- \frac{v_n\{2\}^2}{v_n\{4\}^4 }\frac{\langle v_n^2  V_n V_n^*(p_T)\rangle}{\langle  V_n V_n^*(p_T)\rangle}  \right]   \nonumber\\
\end{eqnarray}
substituting in $2v_n\{2\}^4=v_n\{4\}^4+\langle v_n^4\rangle$ 
\begin{eqnarray}
\frac{v_n\{4\}(p_T) }{v_n\{2\}(p_T) }&=&\frac{v_n\{4\} }{v_n\{2\} } \left[1+\frac{\langle v_n^4\rangle}{v_n\{4\}^4 }- \frac{v_n\{2\}^2}{v_n\{4\}^4 }\frac{\langle v^2  V V^*(p_T)\rangle}{\langle V V^*(p_T)\rangle} \right]   \nonumber\\
&=&\frac{v_n\{4\} }{v_n\{2\} } \left[1+\left(\frac{v_n\{2\}}{v_n\{4\}}\right)^4\left(\frac{\langle  v_n^4\rangle}{\langle  v_n^2\rangle^2}- \frac{\langle v_n^2  V_n V_n^*(p_T)\rangle}{\langle v_n^2\rangle\langle  V_n V_n^*(p_T)\rangle}\right)  \right]\nonumber\\   \label{eqn:corterm}
\end{eqnarray}
If the soft and hard flow harmonics fluctuated in the exact same manner then the magnitude of the elliptical flow across all $p_T$, $|V_n(p_T)|$, would be the magnitude  of the integrated elliptical flow $|V_n|$ multiplied with a function that is only depended on $p_T$:  $|V_n(p_T)|\sim \chi_2(p_T) |V_n|$, which means that $\frac{\langle  v_n^4\rangle}{\langle  v_n^2\rangle^2}- \frac{\langle v_n^2  V_n V_n^*(p_T)\rangle}{\langle v_n^2\rangle\langle  V_n V_n^*(p_T)\rangle}\rightarrow 0$ and then the ratio $\frac{v_n\{4\} }{v_n\{2\} }(p_T)=\frac{v_n\{4\} }{v_n\{2\} }$ would be constant across $p_T$.

Thus, the deviation from $\frac{v_n\{4\} }{v_n\{2\} }(p_T)=\frac{v_n\{4\} }{v_n\{2\} }$ in Figs.\ \ref{fig:v2v4ratlo}-\ref{fig:v2v4rathi} implies that the relationship between the integrated elliptical flow is not linear with the differential elliptical flow.  Indeed, the correction term to $\frac{v_n\{4\} }{v_n\{2\} }$ in Eq.\ (\ref{eqn:corterm}) returns the exact deviation seen in Figs.\ \ref{fig:v2v4ratlo}-\ref{fig:v2v4rathi} and is typically between $0.02-0.06$ with the exception of centrality classes in the $0-10\%$ range.

Experimentally, it is possible to determine the difference between the soft and hard fluctuations, which we defined as $\Delta^{SH}_n$ in \eqref{eqn:deltash}
\begin{eqnarray}
\Delta^{SH}_n(p_T)&\equiv & \underbrace{\frac{\langle  v_n^4\rangle}{\langle  v_n^2\rangle^2}}_{\text{soft fluctuations}}- \underbrace{\frac{\langle v_n^2  V_n V_n^*(p_T)\rangle}{\langle v_n^2\rangle\langle  V_n V_n^*(p_T)\rangle}}_{\text{hard fluctuations}}\nonumber\\
&=& \underbrace{\left(\frac{v_n\{2\}}{v_n\{4\}}\right)^5\left[ \frac{v_n\{4\}(p_T) }{v_n\{2\}(p_T) }-\frac{v_n\{4\} }{v_n\{2\} }\right]}_{\text{Experimental observable}}
\end{eqnarray}
where we rearranged  Eq.\ (\ref{eqn:corterm}) to obtain the experimental observable.  Note that the definition of $\Delta^{SH}_n$  is exact when no multiplicity weighing is used to recombine centrality bins.

\end{document}